\documentclass[journal,10pt,twoside]{IEEEtran} 
\pdfoutput=1
\usepackage{amsmath,cite,amsfonts,amssymb,psfrag,amsthm,paralist}
\usepackage{graphicx}
\usepackage{epstopdf}
\usepackage{rotating}
\usepackage{dsfont}
\usepackage{color}
\usepackage{tikz}
\usetikzlibrary{plotmarks}
\usepackage{pgfplots}
\usetikzlibrary{calc}
\usetikzlibrary{shapes,arrows}
\usetikzlibrary{decorations.markings}
\usetikzlibrary{positioning}
\pgfplotsset{compat=1.10}
\usetikzlibrary{calc}
\usetikzlibrary{shapes,arrows}
\usetikzlibrary{decorations.markings}

\usepackage[utf8]{inputenc}
\usepackage{authblk}
\usepackage{balance}

\DeclareFontFamily{U}{mathx}{\hyphenchar\font45}
\DeclareFontShape{U}{mathx}{m}{n}{<-> mathx10}{}
\DeclareSymbolFont{mathx}{U}{mathx}{m}{n}
\DeclareMathAccent{\widebar}{0}{mathx}{"73}

\newcommand{\Enc}{\mathsf{Enc}}
\newcommand{\Dec}{\mathsf{Dec}}

\makeatletter
\newtheorem*{rep@theorem}{\rep@title}
\newcommand{\newreptheorem}[2]{%
	\newenvironment{rep#1}[1]{%
		\def\rep@title{\Cref{##1}}%
		\begin{rep@theorem}}%
		{\end{rep@theorem}}}
\newcommand*{\textlabel}[2]{%
	\edef\@currentlabel{#1}
	\phantomsection
	#1\label{#2}
}
\makeatother

\newtheorem{theorem}{Theorem}

\newtheorem{remark}{Remark}
\newtheorem{definition}{Definition}

\newtheorem{corollary}{Corollary}

\begin{document}
\title{Secure Multi-Function Computation with\\ Private Remote Sources}
\IEEEoverridecommandlockouts

\author{Onur G\"unl\"u,~\IEEEmembership{Member,~IEEE},  Matthieu Bloch,~\IEEEmembership{Senior Member,~IEEE}, and {Rafael~F.~Schaefer},~\IEEEmembership{Senior~Member,~IEEE}
	\thanks{This work has been supported in part by the German Research Foundation (DFG) under the Grant SCHA 1944/9-1 and in part by the National Science Foundation (NSF) under the Grant CCF 1955401.}
	\thanks{O. G\"unl\"u and R. F. Schaefer are with the Chair of Communications Engineering and Security, University of Siegen, 57076 Siegen, Germany (email: \{onur.guenlue, rafael.schaefer\}@uni-siegen.de).}
	\thanks{M. Bloch is with the School of Electrical and Computer Engineering, Georgia Institute of Technology, Atlanta, GA 30332 (email: matthieu.bloch@ece.gatech.edu).}
}

\maketitle

\begin{abstract}

 We consider a distributed function computation problem in which parties observing noisy versions of a remote source facilitate the computation of a function of their observations at a fusion center through public communication. The distributed function computation is subject to constraints, including not only reliability and storage but also privacy and secrecy. Specifically,
  \begin{inparaenum}
  \item the remote source should remain \emph{private} from an eavesdropper and the fusion center, measured in terms of the information leaked about the remote source;
  \item the function computed should remain \emph{secret} from the eavesdropper, measured in terms of the information leaked about the arguments of the function, to ensure secrecy regardless of the exact function used.
  \end{inparaenum}
  We derive the exact rate regions for lossless and lossy single-function computation and illustrate the lossy single-function computation rate region for an information bottleneck example, in which the optimal auxiliary random variables are characterized for binary-input symmetric-output channels. We extend the approach to lossless and lossy asynchronous multiple-function computations with joint secrecy and privacy constraints, in which case inner and outer bounds for the rate regions differing only in the Markov chain conditions imposed are characterized.

\end{abstract}
%
\IEEEpeerreviewmaketitle
\section{Introduction} \label{sec:intro}
Consider a scenario in which multiple terminals that observe dependent random sequences want to compute a function of their sequences by exchanging messages through public communication links \cite{YaoSecureFunctionComp,YaoSecureFunctionComp2}. One application for which this distributed function computation problem is relevant is network function virtualization \cite{NFVReview} via, e.g., software defined networking. To compute a function of a subset of sequences, it is not always necessary for the terminal computing the function, called \emph{fusion center}, to observe the exact sequences \cite{CodingforComputing}. This fact allows one to reduce the public communication rate, also called \emph{storage rate}, required for reliable function computations by using, e.g., distributed lossless  source coding techniques \cite{SW}. Furthermore, for various functions, it suffices to recover a distorted version of the original sequence by using, e.g., distributed lossy source coding methods \cite{WZCard}. Lossy reconstruction allows to further reduce the amount of public storage, which is useful for resource-limited networks such as Internet-of-Things (IoT) devices that make aggregated decisions using lightweight mechanisms \cite{CodingforComputing,bizimEntropyTutorial,csiszarnarayan, bizimWZ,IoTCompute,OurJSAITTutorial}; see \cite{IshwarCompute,TchamkertenCompute,GastparCompute,KumarCompute,ViswanathCompute} for various extensions of the basic function computation problem with reliability and storage constraints. 

Reliable function computation and small public storage constraints have also been combined with \emph{secrecy} constraints, requiring that the computed function outputs be hidden from an eavesdropper \cite{TyagiNarayanSecureCompute}. In addition to the public messages exchanged between terminals, the eavesdropper is considered to have access to a random sequence correlated with other sequences. Various extensions of the basic secure function computation or distributed source coding problems have been analyzed in the literature~\cite{TyagiWatanabeEUROCrypt,TyagiWatanabeTIT,PrabhakaranCompute, GoldenbaumComp, GunduzCompute,PrabhakaranISIT2020, TyagiJSAC2013}. Furthermore, a \emph{privacy} constraint has been added in \cite{LifengFCTrans} to the problem. The main difference between \emph{secrecy} and \emph{privacy} is that secrecy leakage is measured with respect to the functions computed while privacy leakage is measured with respect to the source sequences themselves. A privacy leakage analysis provides an upper bound on the secrecy leakage of future function computations involving the terminals already participating in earlier function computations \cite{benimdissertation,bizimMMMMTIFS}. This is because the information leaked about the sequence of a terminal might leak information about another function computed by using the same sequence. We extend \cite{LifengFCTrans} by considering separate privacy constraints on the source of the random sequence of the \emph{transmitting terminal} that sends a public message to the fusion center. 

A common assumption in the literature is that sequences observed by all terminals are distributed according to a joint probability distribution. However, the correlated random sequences observed by terminals in a network generally stem from a common source of information, e.g., some sensor location information transmitted through the network before the next function computation starts, distorted versions of which are distributed within the network. Thus, we posit that there exists a common true source, called the ground truth or the remote source, hidden from all terminals and of which the observed sequences are noisy versions. Such a hidden source model allows a terminal to combine multiple observed sequences to obtain a single ``higher quality" random sequence, which is entirely similar to applying maximal ratio combining over an additive white Gaussian noise (AWGN) channel. This approach is thus useful to model the quality differences between random sequences observed by different terminals. If the function computation network is mistakenly modeled with a visible (or unhidden) source model, the code construction designed for the assumed visible source model might result in unnoticed secrecy leakage and reduction in computation reliability, as illustrated in \cite{bizimMMMMTIFS} for key agreement. 

Noisy measurements of a hidden source are generally modeled as observations through broadcast channels (BCs) \cite{CoverandThomas} to have a generic measurement model that allows noise components at different terminals to be correlated \cite{bizimITW,CorrelatedPaperLong}. Such a hidden source model is proposed and motivated in \cite{groundtruthauthentication} for authentication problems and in \cite{bizimKittipongTIFS, bizimITW} for secret-key agreement problems with a privacy constraint. As we detail in Section~\ref{sec:problem_setting}, such a hidden source model results in two different privacy leakage constraints measured with respect to the hidden source, which is different from the single privacy leakage constraint considered in \cite{LifengFCTrans} measured with respect to the random sequence observed by the transmitting terminal. The privacy leakage and storage rates are shown below to be different for this model, unlike in previous works. Furthermore, the equivocation of the source is commonly used in the literature to measure the secrecy leakage, which results in rate bounds with conditional entropy terms. By replacing the equivocation with the mutual information terms, we obtain rate regions with simpler notation and easier interpretations.

We consider two function computation settings. The first setting imposes a reliable (\textit{lossless}) computation of the function of interest and the other one allows a fixed level of distortion between the computed function and the actual function output (\textit{lossy function computation}) \cite{LifengFCTrans}. These settings address different applications. For instance, the lossless function computation setting might model user/terminal identification, where the exact identifier recovery is necessary; in contrast, the lossy function computation setting might model user/terminal authentication, where a set of users whose computed functions are close to a pre-defined value are authenticated. We bound the error probability for the reliable function computation task for the lossless setting and the expected distortion for lossy setting, respectively, which require different proof steps. We exactly characterize the rate regions for both settings when a single function is computed. 

We further extend the function computation with privacy and secrecy problem by considering multiple function computations with \textit{joint secrecy and privacy constraints} on all terminals involved in any function computation task. This extension allows one to measure the total amount of information about all computed functions within a network leaked to an eavesdropper. This extension also allows one to correctly characterize the privacy leakage to an eavesdropper, i.e., the amount of information about the hidden source leaked to an eavesdropper who might observe all public messages and all side information obtained during all (not necessarily synchronous) function computations within the same network. Multiple function computations with joint secrecy and privacy constraints are closely related to the multi-entity and multi-enrollment key agreement problems in \cite{benimmultientityTIFS}, where the noisy measurements of the same hidden source are used for multiple key agreements. Both lossless and lossy function computation settings are analyzed to provide inner and outer bounds for the multi-function rate regions, for which only the imposed Markov chains differ.

\subsection{Summary of Contributions}
Our problem formulation introduces one secrecy and two privacy constraints, in addition to reliability (or distortion) and storage constraints, to the single function computation problem to characterize the resulting rate regions. These results are strict extensions of \cite{LifengFCTrans} as we consider a remote source common to all terminals with side information sequences that are noisy measurements of the remote source. Furthermore, we also consider multiple asynchronous function computations within the same network with joint secrecy and privacy constraints over all terminals involved in any function computation. A summary of the main contributions is as follows.

\begin{itemize}
	\item We derive the rate region for lossless single-function computation with secrecy and privacy constraints. The remote source model we consider corresponds to a physically-degraded BC and when the transmitting observes the remote (noiseless) source outputs, the model reduces to a semi-deterministic BC. Furthermore, we show that convexification with a time-sharing random variable is necessary, which is missing in some previous works.
	\item We next consider the lossless multi-function computation where a finite number $J$ of functions are computed from different noisy measurements (observed by different terminals) of the same remote source asynchronously. We impose one secrecy and privacy constraints that consider the total leakage in the network, i.e., they are joint constraints for all parties involved in any function computation. We propose inner and outer bounds for the multi-function rate region that differ only in the Markov chain conditions imposed on the auxiliary random variables. The rate regions include both separate constraints for each terminal and joint constraints for all terminals. 
	\item All inner and outer bounds for the lossless single- and multi-function computations are extended to the corresponding lossy settings. Similar to the lossless case, we characterize the lossy rate region for the single-function computation, and we provide inner and outer bounds for the multi-function computations that differ only in the Markov chains imposed.  
	\item We evaluate the rate region for a lossy single-function computation problem, in which the measurement channel of the eavesdropper is physically-degraded compared to the measurement channel of the fusion center. We solve an information bottleneck problem to obtain the rate region boundary tuples.
\end{itemize}

\subsection{Organization}
This paper is organized as follows. In Section~\ref{sec:problem_setting}, we describe four function computation problems with a remote source that are lossless or lossy and single-function or multi-function computation problems. We present the rate regions for the lossless and lossy single-function computation in Section~\ref{sec:rateregions} in addition to inner and outer bounds with different Markov chains for the lossless and lossy multi-function computations for any finite number of functions. In Section~\ref{sec:example}, we solve an information bottleneck problem to illustrate the rate region for the lossy single-function computation problem. In Section~\ref{sec:proofTheorem1}, we provide the detailed proof for characterizing the rate regions of the lossless single-function computation. Similarly, we offer proofs of the inner and outer bounds for the lossless multi-function computations in Section~\ref{sec:proofTheorem3}. In Section~\ref{sec:conclusion}, we conclude the paper.

\subsection{Notation}
Upper case letters represent random variables and lower case letters their realizations. A superscript denotes a sequence of variables, e.g., $\displaystyle X^n\!=\!X_1,X_2,\ldots, X_i,\ldots, X_n$, and a subscript $i$ denotes the position of a variable in a sequence. A random variable $\displaystyle X$ has probability distribution $\displaystyle P_X$. Calligraphic letters such as $\displaystyle \mathcal{X}$ denote sets, set sizes are written as $\displaystyle |\mathcal{X}|$ and their complements as $\displaystyle \mathcal{X}^c$. $[1\!:\!J]$ denotes the set $\{1,2,\ldots,J\}$ for an integer $J\geq1$ and $[1\!:\!J]\!\setminus\!\{j\}$ denotes the set $\{1,2,\ldots,j-1,j+1,\ldots,J\}$ for any $j\in[1\!:\!J]$. $H_b(x)\!=\!-x\log x- (1\!-\!x)\log (1\!-\!x)$ is the binary entropy function, where logarithms are to the base $2$, and $H_b^{-1}(\cdot)$ denotes its inverse with range $[0, 0.5]$. A binary symmetric channel (BSC) with crossover probability $p$ is denoted by BSC($p$). $X\sim\text{Bern}(\alpha)$ is a binary random variable with $\Pr[X=1]=\alpha$.

\section{Problem Definitions}\label{sec:problem_setting}

\subsection{Lossless Single-Function Computation}\label{subsec:singlefunction}
\begin{figure}
	\centering
	\resizebox{0.97\linewidth}{!}{
		\begin{tikzpicture}
			\node (so) at (-1.5,-3.3) [draw,rounded corners = 5pt, minimum width=0.8cm,minimum height=0.8cm, align=left] {$P_X$};
			\node (a) at (0,-0.5) [draw,rounded corners = 6pt, minimum width=2.2cm,minimum height=0.8cm, align=left] {$W = \Enc(\widetilde{X}^n)$};
			\node (c) at (5,-3.3) [draw,rounded corners = 5pt, minimum width=1.3cm,minimum height=0.6cm, align=left] {$P_{YZ|X}$};
			\node (f) at (0,-2.05) [draw,rounded corners = 5pt, minimum width=1cm,minimum height=0.6cm, align=left] {$P_{\widetilde{X}|X}$};
			\node (b) at (5,-0.5) [draw,rounded corners = 6pt, minimum width=2.2cm,minimum height=0.8cm, align=left] {$\widehat{f^n}= \Dec\left(W,Y^n\right)$};
			\node (g) at (5,-5) [draw,rounded corners = 5pt, minimum width=1cm,minimum height=0.6cm, align=left] {Eve};
			\draw[decoration={markings,mark=at position 1 with {\arrow[scale=1.5]{latex}}},
			postaction={decorate}, thick, shorten >=1.4pt] (a.east) -- (b.west) node [midway, above] {$W$};
			\node (a1) [below of = a, node distance = 2.8cm] {$X^n$};
			\draw[decoration={markings,mark=at position 1 with {\arrow[scale=1.5]{latex}}},
			postaction={decorate}, thick, shorten >=1.4pt] ($(c.north)+(0.0,0)$) -- ($(b.south)+(0.0,0)$) node [midway, left] {$Y^n$};
			\draw[decoration={markings,mark=at position 1 with {\arrow[scale=1.5]{latex}}},
			postaction={decorate}, thick, shorten >=1.4pt] (so.east) -- (a1.west);
			\draw[decoration={markings,mark=at position 1 with {\arrow[scale=1.5]{latex}}},
			postaction={decorate}, thick, shorten >=1.4pt] (a1.north) -- (f.south);
			\draw[decoration={markings,mark=at position 1 with {\arrow[scale=1.5]{latex}}},
			postaction={decorate}, thick, shorten >=1.4pt] (f.north) -- (a.south) node [midway, right] {$\widetilde{X}^n$};
			\draw[decoration={markings,mark=at position 1 with {\arrow[scale=1.5]{latex}}},
			postaction={decorate}, thick, shorten >=1.4pt, dashed] (a1.east) -- ($(c.west)-(0,0.0)$) node [above  left] {$X^n$};
			\draw[decoration={markings,mark=at position 1 with {\arrow[scale=1.5]{latex}}},
			postaction={decorate}, thick, shorten >=1.4pt] (c.south) -- (g.north) node [midway, right] {$Z^n$};
			\node (b2) [right of = b, node distance = 2.5cm] {$\widehat{f^n}$};
			\draw[decoration={markings,mark=at position 1 with {\arrow[scale=1.5]{latex}}},
			postaction={decorate}, thick, shorten >=1.4pt] (b.east) -- (b2.west);
			\draw[decoration={markings,mark=at position 1 with {\arrow[scale=1.5]{latex}}},
			postaction={decorate}, thick, shorten >=1.4pt] ($(a.east)+(1,0)$) -- ($(a.east)+(1,-4.5)$) -- ($(a.east)+(1,-4.5)$) -- (g.west) node [above left=0.0cm and 0.5cm of g.west] {$W$};
		\end{tikzpicture}
	}
	\caption{Noisy measurements of a remote source used to compute a function securely and privately with the help of a public communication link.}\label{fig:hidden}
\end{figure}
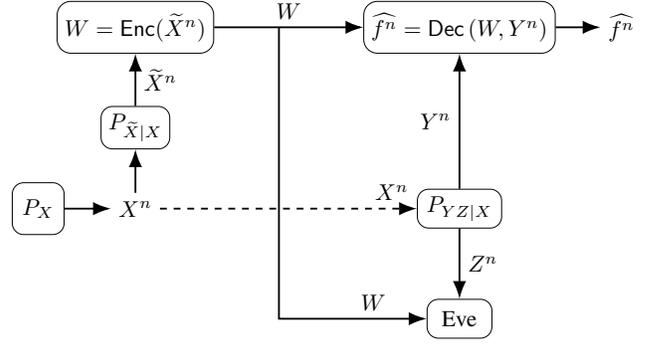

Consider the function computation model illustrated in Fig.~\ref{fig:hidden}. Three terminals obtain noisy observations $\widetilde{X}^n,Y^n,Z^n$, respectively, of a single  i.i.d. remote source $X^n$, through a memoryless channel with transition probability $p_{\widetilde{X}|X}p_{YZ|X}$. The source alphabet $\mathcal{X}$ and measurement alphabets $\widetilde{\mathcal{X}},\mathcal{Y}, \mathcal{Z}$ are finite sets. The objective is for the terminal observing $\widetilde{X}^n$ to transmit a message $W= \Enc(\widetilde{X}^n)$ over a public channel and to enable the terminal observing $Y^n$ to compute a function $f^n(\widetilde{X}^n,Y^n)$ such that 
\begin{align}
	f^n(\widetilde{X}^n,Y^n) = {\{f(\widetilde{X}_i,Y_i)\}}_{i=1}^n.
\end{align}
The terminal observing $Z^n$ and obtaining $W$ through the public channel is treated as an eavesdropper (Eve).

Since $P_{\widetilde{X}XYZ}$ is fixed, the separate measurement channels $P_{\widetilde{X}|X}$ and $P_{YZ|X}$ in Fig.~\ref{fig:hidden} can be modeled as a physically-degraded BC with transition probability $P_{XYZ|\widetilde{X}}=P_{X|\widetilde{X}}P_{YZ|X}$ and with fixed input probability distribution $P_{\widetilde{X}}$. For such a BC, the case of a noiseless measurement for which $\widetilde{X}^n=X^n$ can be treated as a semi-deterministic BC. 

\begin{definition}
	\normalfont A tuple $(R_{\text{s}}, R_{\text{w}},R_{\ell,{\text{Dec}}}, R_{\ell,{\text{Eve}}})$ is \emph{achievable} if, for any $\delta\!>\!0$, there exist $n\!\geq\!1$, an encoder, and a decoder such that
	\begin{align}
	&\Pr\Big[f^n(\widetilde{X}^n,Y^n) \neq \widehat{f^n}\Big] \leq \delta&& (\text{reliability})\label{eq:reliability_cons}\\
	& \frac{1}{n}I(\widetilde{X}^n,Y^n;W|Z^n) \leq R_{\text{s}}+\delta&&(\text{secrecy})\label{eq:secrecyleakage_cons}\\
		&\frac{1}{n}\log\big|\mathcal{W}\big| \leq R_{\text{w}}+\delta&&(\text{storage})\label{eq:storage_cons}\\
	&\frac{1}{n}I(X^n;W|Y^n) \leq R_{\ell,\text{Dec}}+\delta&&(\text{privacyDec})\label{eq:privDec_cons}\\
	&\frac{1}{n}I(X^n;W|Z^n) \leq R_{\ell,\text{Eve}}+\delta&&(\text{privacyEve})\label{eq:privEve_cons}.
	\end{align}
    The region $\mathcal{R}$ is the closure of the set of all achievable tuples.\hfill $\lozenge$
\end{definition}

Note that the metric $I(f^n(\widetilde{X}^n,Y^n); W|Z^n)$ might seem  a more natural way to measure the information leakage to the eavesdropper who observes $(W,Z^n)$ of the computed function $f^n(\cdot,\cdot)$. However, the analysis of this metric depends on the specific properties of the function $f(\cdot,\cdot)$. Since the data-processing inequality ensures that $I(f^n(\widetilde{X}^n,Y^n); W|Z^n)\leq I(\widetilde{X}^n,Y^n;W|Z^n)$ for all functions $f(\cdot,\cdot)$ with equality if $f(\cdot,\cdot)$ is a bijective mapping, we instead consider the metric in (\ref{eq:secrecyleakage_cons}). The analysis then does not depend on the computed function $f(\cdot,\cdot)$ and provides a valid upper bound on the proper secrecy-leakage rate metric for any $f(\cdot,\cdot)$. Since $I(\widetilde{X}^n,Y^n;W|Z^n)=I(\widetilde{X}^n;W|Z^n)$ because of the Markov chain $W-\widetilde{X}^n-(Y^n,Z^n)$, the equivocation $H(\widetilde{X}^n|W,Z^n)$ considered in previous works \cite{LifengFCTrans} captures the same secrecy leakage as (\ref{eq:secrecyleakage_cons}). Furthermore, the privacy leakage metrics in (\ref{eq:privDec_cons}) and (\ref{eq:privEve_cons}) measure the information leakage about the remote source to the decoder and eavesdropper, respectively, due to function computation. We remark that in (\ref{eq:secrecyleakage_cons}), (\ref{eq:privDec_cons}), and (\ref{eq:privEve_cons}), we consider conditional mutual information terms to take into consideration the unavoidable secrecy or privacy leakage due to side information available at the fusion center or eavesdropper.

\subsection{Lossy Single-Function Computation}\label{subsec:singlefunctionlossy}
Consider again the single-function computation model depicted in Fig.~\ref{fig:hidden} and replace the reliability constraint in (\ref{eq:reliability_cons}) with an expected distortion constraint to allow a distorted reconstruction of the function $f(\cdot,\cdot)$. This defines the lossy single-function computation model, for which the notion of achievability is as follows.

\begin{definition}
	\normalfont A \emph{lossy} tuple $(R_{\text{s}}, R_{\text{w}},R_{\ell,{\text{Dec}}}, R_{\ell,{\text{Eve}}},D)$ is \emph{achievable} if, for any $\delta\!>\!0$, there exist $n\!\geq\!1$, an encoder, and a decoder that satisfy (\ref{eq:secrecyleakage_cons})-(\ref{eq:privEve_cons}) and
	\begin{align}
	&\mathbb{E}\Big[d(f^n(\widetilde{X}^n,Y^n),\widehat{f^n})\Big] \leq D+\epsilon&&\quad \label{eq:reliability_conslossysingle}
	\end{align}
	where $d(f^n,\widehat{f^n})=\frac{1}{n}\sum_{i=1}^nd(f_i,\widehat{f}_i)$ is a per-letter distortion metric. The \emph{lossy} region $\mathcal{R}_{\text{D}}$ is the closure of the set of all achievable lossy distortion tuples.\hfill $\lozenge$
\end{definition}

\subsection{Lossless Multi-Function Computation}\label{subsec:losslesmultifunction}
\begin{figure}
	\centering
	\resizebox{0.97\linewidth}{!}{
		\begin{tikzpicture}
			\node (so) at (-1.5,-3.3) [draw,rounded corners = 5pt, minimum width=0.8cm,minimum height=0.8cm, align=left] {$P_X$};
			\node (a) at (0,-0.5) [draw,rounded corners = 6pt, minimum width=2.2cm,minimum height=0.8cm, align=left] {$W_1 = \Enc_1(\widetilde{X}_1^n)$};
			\node (c) at (5,-3.3) [draw,rounded corners = 5pt, minimum width=1.3cm,minimum height=0.6cm, align=left] {$P_{Y_1Z_1|X}$};
			\node (f) at (0,-2.05) [draw,rounded corners = 5pt, minimum width=1cm,minimum height=0.6cm, align=left] {$P_{\widetilde{X}_1|X}$};
			\node (b) at (5,-0.5) [draw,rounded corners = 6pt, minimum width=2.2cm,minimum height=0.8cm, align=left] {$\widehat{f_1^n}= \Dec_1\left(W_1,Y_1^n\right)$};
			\node (g) at (5,-5) [draw,rounded corners = 5pt, minimum width=1cm,minimum height=0.6cm, align=left] {Eve};
			\draw[decoration={markings,mark=at position 1 with {\arrow[scale=1.5]{latex}}},
			postaction={decorate}, thick, shorten >=1.4pt] (a.east) -- (b.west) node [midway, above] {$W_1$};
			\node (a1) [below of = a, node distance = 2.8cm] {$X^n$};
			\draw[decoration={markings,mark=at position 1 with {\arrow[scale=1.5]{latex}}},
			postaction={decorate}, thick, shorten >=1.4pt] ($(c.north)+(0.0,0)$) -- ($(b.south)+(0.0,0)$) node [midway, right] {$Y_1^n$};
			\draw[decoration={markings,mark=at position 1 with {\arrow[scale=1.5]{latex}}},
			postaction={decorate}, thick, shorten >=1.4pt] (so.east) -- (a1.west);
			\draw[decoration={markings,mark=at position 1 with {\arrow[scale=1.5]{latex}}},
			postaction={decorate}, thick, shorten >=1.4pt] (a1.north) -- (f.south);
			\draw[decoration={markings,mark=at position 1 with {\arrow[scale=1.5]{latex}}},
			postaction={decorate}, thick, shorten >=1.4pt] (f.north) -- (a.south) node [midway, left] {$\widetilde{X}_1^n$};
			\draw[decoration={markings,mark=at position 1 with {\arrow[scale=1.5]{latex}}},
			postaction={decorate}, thick, shorten >=1.4pt, dashed] (a1.east) -- ($(c.west)-(0,0.0)$) node [above  left] {$X^n$};
			\draw[decoration={markings,mark=at position 1 with {\arrow[scale=1.5]{latex}}},
			postaction={decorate}, thick, shorten >=1.4pt] (c.south) -- (g.north) node [midway, right] {$Z_1^n$};
			\node (b2) [right of = b, node distance = 2.5cm] {$\widehat{f_1^n}$};
			\draw[decoration={markings,mark=at position 1 with {\arrow[scale=1.5]{latex}}},
			postaction={decorate}, thick, shorten >=1.4pt] (b.east) -- (b2.west);
			\draw[decoration={markings,mark=at position 1 with {\arrow[scale=1.5]{latex}}},
			postaction={decorate}, thick, shorten >=1.4pt] ($(a.east)+(1,0)$) -- ($(a.east)+(1,-4.35)$) -- ($(a.east)+(1,-4.35)$) -- ($(g.west)+(0,0.15)$) node [above left=0.15cm and 0.5cm of g.west] {$W_1$};
			\node (a2) at (0,-8.5) [draw,rounded corners = 6pt, minimum width=2.2cm,minimum height=0.8cm, align=left] {$W_2 = \Enc_2(\widetilde{X}_2^n)$};
			\node (c2) at (5,-7) [draw,rounded corners = 5pt, minimum width=1.3cm,minimum height=0.6cm, align=left] {$P_{Y_2Z_2|X}$};
			\node (f2) at (0,-7) [draw,rounded corners = 5pt, minimum width=1cm,minimum height=0.6cm, align=left] {$P_{\widetilde{X}_2|X}$};
			\node (b2) at (5,-8.5) [draw,rounded corners = 6pt, minimum width=2.2cm,minimum height=0.8cm, align=left] {$\widehat{f_2^n}= \Dec_2\left(W_2,Y_2^n\right)$};
			\draw[decoration={markings,mark=at position 1 with {\arrow[scale=1.5]{latex}}},
			postaction={decorate}, thick, shorten >=1.4pt] (a2.east) -- (b2.west) node [midway, below] {$W_2$};
			\draw[decoration={markings,mark=at position 1 with {\arrow[scale=1.5]{latex}}},
			postaction={decorate}, thick, shorten >=1.4pt] ($(c2.south)+(0.0,0)$) -- ($(b2.north)+(0.0,0)$) node [midway, right] {$Y_2^n$};
			\draw[decoration={markings,mark=at position 1 with {\arrow[scale=1.5]{latex}}},
			postaction={decorate}, thick, shorten >=1.4pt] (a1.south) -- (f2.north);
			\draw[decoration={markings,mark=at position 1 with {\arrow[scale=1.5]{latex}}},
			postaction={decorate}, thick, shorten >=1.4pt] (f2.south) -- (a2.north) node [midway, left] {$\widetilde{X}_2^n$};
			\draw[decoration={markings,mark=at position 1 with {\arrow[scale=1.5]{latex}}},
			postaction={decorate}, thick, shorten >=1.4pt, dashed] ($(a1.east) + (1,0)$) -- ($(a1.east) + (1,-3.7)$) -- ($(a1.east) + (1,-3.7)$) -- ($(c2.west)-(0,0.0)$) node [below  left] {$X^n$};
			\draw[decoration={markings,mark=at position 1 with {\arrow[scale=1.5]{latex}}},
			postaction={decorate}, thick, shorten >=1.4pt] (c2.north) -- (g.south) node [midway, right] {$Z_2^n$};
			\node (b22) [right of = b2, node distance = 2.5cm] {$\widehat{f_2^n}$};
			\draw[decoration={markings,mark=at position 1 with {\arrow[scale=1.5]{latex}}},
			postaction={decorate}, thick, shorten >=1.4pt] (b2.east) -- (b22.west);
			\draw[decoration={markings,mark=at position 1 with {\arrow[scale=1.5]{latex}}},
			postaction={decorate}, thick, shorten >=1.4pt] ($(a2.east)+(1,0)$) -- ($(a2.east)+(1,3.35)$) -- ($(a2.east)+(1,3.35)$) -- ($(g.west)+(0,-0.15)$) node [below left=0.15cm and 0.5cm of g.west] {$W_2$};
		\end{tikzpicture}
	}
	\caption{Noisy measurements of the same remote source used to compute $J=2$ functions (via $2J = 4$ parties) securely and privately with the help of public communication links.}\label{fig:hiddenmultifunctions}
\end{figure}
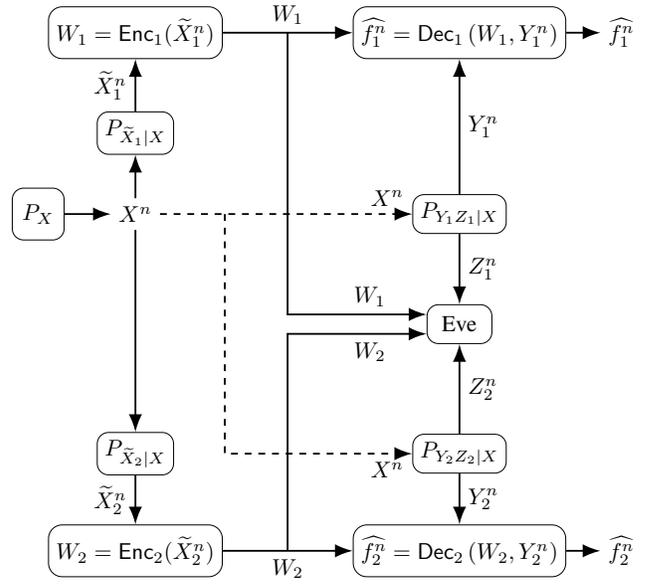

We next extend the lossless single-function computation model by considering that the same remote source $X^n$ is measured by multiple encoder and decoder pairs to compute different functions. Consider a finite number $J\geq 1$ of encoders $\Enc_j(\widetilde{X}_j)=W_j$, decoders $\Dec_j(W_j,Y_j^n)=\widehat{f_j^n}$, and functions $f_j^n(\widetilde{X}_j^n,Y_j^n)=\{f_j(\widetilde{X}_{i,j},Y_{i,j})\}_{i=1}^n$ for $j\in[1:J]$, where $\widetilde{X}_j^n$ is measured through the channel $P_{\widetilde{X}_j|X}$ and $(Y_j^n,Z_j^n)$ are measured through the BC $P_{Y_jZ_j|X}$. The eavesdropper observes $(Z_{[1:J]}^n, W_{[1:J]})$. This multi-function computation model is illustrated in Fig.~\ref{fig:hiddenmultifunctions} for $J=2$.

\begin{definition}\label{def:multifunctioncomputation}
	\normalfont A \emph{multi-function} tuple $(R_\text{s}, R_{\text{w},[1:J]},R_{\ell,{\text{Dec}},[1:J]}, R_{\ell,{\text{Eve}}})$ with $j$-th encoder measurements through $P_{\widetilde{X}_j|X}$ and $j$-th decoder measurements through $P_{Y_jZ_j|X}$ for all $j\in[1:J]$ is \emph{achievable} if, for any $\delta\!>\!0$, there exist $n\!\geq\!1$, and $J$ encoder and decoder pairs such that
	\begin{alignat}{2}
		&\Pr\Bigg[\underset{j\in[1:J]}{\bigcup}\Big\{ f_j^n(\widetilde{X}_j^n,Y^n_j)\neq \widehat{f_j^n}\Big\}\Bigg] \leq \delta&&\label{eq:reliability_consmfc}\\
		& \!\frac{1}{n}I(\widetilde{X}_{[1:J]}^n,Y_{[1:J]}^n;W_{[1:J]}|Z_{[1:J]}^n\!) \!\leq\! R_{\text{s}}\!+\!\delta&&\label{eq:secrecyleakage_consmfc}\\
		&\frac{1}{n}\log\big|\mathcal{W}_j\big| \leq R_{\text{w},j}+\delta,&&\forall j\in[1:J]\label{eq:storage_consmfc}\\
		&\frac{1}{n}I(X^n;W_j|Y_j^n) \leq R_{\ell,\text{Dec},j}+\delta,&&\forall j\in[1:J]\label{eq:privDec_consmfc}\\
		&\frac{1}{n}I(X^n;W_{[1:J]}|Z_{[1:J]}^n) \leq R_{\ell,\text{Eve}}+\delta.&&\label{eq:privEve_consmfc}
	\end{alignat}
	The \emph{multi-function} region $\mathcal{R}_{\text{mf}}$ is the closure of the set of all achievable tuples.\hfill $\lozenge$
\end{definition}

\begin{remark}
	The storage rate constraints in (\ref{eq:storage_consmfc}) and the corresponding privacy leakage rate constraints in (\ref{eq:privDec_consmfc}) are $J$ separate constraints. However, the reliability constraint in (\ref{eq:reliability_consmfc}), the secrecy leakage constraint in (\ref{eq:secrecyleakage_consmfc}), and the privacy leakage rate constraint in (\ref{eq:privEve_consmfc}) are joint constraints that depend on the parameters of all $J$ encoder-decoder pairs.
\end{remark}

\subsection{Lossy Multi-Function Computation}\label{subsec:multiplefunctionlossy}
Similar to Section~\ref{subsec:singlefunctionlossy}, we extend the model of Section~\ref{subsec:losslesmultifunction} to allow distorted function computations for multiple functions $f_j^n(\widetilde{X}_j^n,Y_j^n)=\{f_j(\widetilde{X}_{i,j},Y_{i,j})\}_{i=1}^n$ computed from different measurements $(\widetilde{X}^n_j,Y^n_j)$ of the same remote source $X^n$.

\begin{definition}
	\normalfont A \emph{lossy multi-function} tuple $(R_{\text{s}}, R_{\text{w},[1:J]},R_{\ell,{\text{Dec}},[1:J]}, R_{\ell,{\text{Eve}}},D_{[1:J]})$ with $j$-th encoder measurements through $P_{\widetilde{X}_j|X}$ and $j$-th decoder measurements through $P_{Y_jZ_j|X}$ for all $j\in[1:J]$ is \emph{achievable} if, for any $\delta\!>\!0$, there exist $n\!\geq\!1$, and $J$ encoder and decoder pairs that satisfy (\ref{eq:secrecyleakage_consmfc})-(\ref{eq:privEve_consmfc}) and
	\begin{align}
	&\mathbb{E}\Big[d(f_j^n(\widetilde{X}^n_j,Y_j^n),\widehat{f_j^n})\Big] \leq D_j+\delta,\qquad\quad\forall j\in[1:J]\label{eq:reliability_conslossymultiple}
	\end{align}
	where $d(f^n,\widehat{f^n})=\frac{1}{n}\sum_{i=1}^nd(f_i,\widehat{f}_i)$ is a per-letter distortion metric. The \emph{lossy multi-function} region $\mathcal{R}_{\text{mf,D}}$ is the closure of the set of all achievable lossy distortion tuples.\hfill $\lozenge$
\end{definition}

\section{Rate Regions}\label{sec:rateregions}
We first recall the notion of an \textit{admissible random variable}, used in Theorems~\ref{theo:innerouter_secrecystorprivregions} and \ref{theo:innerouterlosslesmfc}.

\begin{definition}[\hspace{1sp}\cite{CodingforComputing}]
	\normalfont A (vector) random variable $U$ is admissible for a function $f(\widetilde{X},Y)$ if  $U-\widetilde{X}-Y$ form a Markov chain and $H(f(\widetilde{X},Y)|U,Y) = 0$, i.e.,  $(U,Y)$ determine  $f(\widetilde{X},Y)$.\hfill $\lozenge$
\end{definition}

Define $[a]^-=\min\{a,0\}$ and $[a]^+=\max\{a,0\}$ for $a\in\mathbb{R}$.

\subsection{Lossless Single-Function Computation}
We characterize the region $\mathcal{R}$ for the lossless single function computation problem in Theorem~\ref{theo:innerouter_secrecystorprivregions}. The corresponding proof is detailed in Section~\ref{sec:proofofTheorem1}.

\begin{theorem}\label{theo:innerouter_secrecystorprivregions}
	The region $\mathcal{R}$ is the set of all tuples $(R_{\text{s}}, R_{\text{w}},R_{\ell,{\text{Dec}}}, R_{\ell,{\text{Eve}}})$ satisfying
	\begin{align}
	&R_{\text{s}}\!\geq  I(U;\widetilde{X}|Z)+[I(U;Z|V,Q)-I(U;Y|V,Q)]^-\label{eq:secrecyinner}\\
	&R_{\text{w}}\!\geq\! I(U;\widetilde{X}|Y)\label{eq:storageinner}\\
	&R_{\ell,\text{Dec}}\!\geq	\! I(U;X|Y)\label{eq:privacyDecinner}\\
	&R_{\ell,\text{Eve}}\!\geq\! I(U;X|Z)\!+\![I(U;Z|V,Q)\!-\!I(U;Y|V,Q)]^-	\label{eq:privacyEveinner}
	\end{align}
	such that $U$ is admissible for the function $f(\widetilde{X},Y)$ and $(Q,V)-U-\widetilde{X}-X-(Y,Z)$ form a Markov chain. The region $\mathcal{R}$ is convexified by using the time-sharing random variable $Q$, which is required because of the $[\cdot]^-$ operation. One can limit the cardinalities of $Q$, $V$, and $U$ to $|\mathcal{Q}|\leq 2$, $|\mathcal{V}|\leq |\widetilde{X}|+4$, and $|\mathcal{U}|\leq (|\widetilde{X}|+4)^2$.
\end{theorem}

In \cite{LifengFCTrans}, some lower bounds on the rates in the rate regions include terms with the maximization operator $[\cdot]^+$. One can show that the rate regions in \cite{LifengFCTrans} that include such lower bounds are not convex and can be enlarged by using a time-sharing random variable $Q$, as considered in this work. 

\subsection{Lossy Single-Function Computation}
We next characterize the lossy region $\mathcal{R}_D$ for the lossy single function computation problem in Theorem~\ref{theo:innerouter_secrecystorprivregionslossy}.

\begin{theorem}\label{theo:innerouter_secrecystorprivregionslossy}
	The \emph{lossy} region $\mathcal{R}_D$ is the set of all tuples $(R_{\text{s}}, R_{\text{w}},R_{\ell,{\text{Dec}}}, R_{\ell,{\text{Eve}}},D)$ satisfying
	\begin{align}
	&R_{\text{s}}\!\geq  I(U;\widetilde{X}|Z)+[I(U;Z|V,Q)-I(U;Y|V,Q)]^-\label{eq:secrecyinnerlossy}\\
	&R_{\text{w}}\!\geq\! I(U;\widetilde{X}|Y)\label{eq:storageinnerlossy}\\
	&R_{\ell,\text{Dec}}\!\geq	\! I(U;X|Y)\label{eq:privacyDecinnerlossy}\\
	&R_{\ell,\text{Eve}}\!\geq\! I(U;X|Z)\!+\![I(U;Z|V,Q)\!-\!I(U;Y|V,Q)]^-	\label{eq:privacyEveinnerlossy}\\
	&D\geq \mathbb{E}[d(f(\widetilde{X},Y),g(U,Y))]\label{eq:distortioninnerlossy}
	\end{align}
	for some function $g(\cdot,\cdot)$ such that $(Q,V)-U-\widetilde{X}-X-(Y,Z)$ form a Markov chain. One can limit the cardinalities to $|\mathcal{Q}|\leq 2$, $|\mathcal{V}|\leq |\widetilde{X}|+5$, and $|\mathcal{U}|\leq (|\widetilde{X}|+5)^2$.
\end{theorem}

\begin{IEEEproof}[Proof Sketch]
	The achievability proof of Theorem~\ref{theo:innerouter_secrecystorprivregionslossy} follows from the achievability proof of Theorem~\ref{theo:innerouter_secrecystorprivregions}, except that $U$ is not necessarily admissible, and with the addition that $P_{U|\widetilde{X}}$ and $P_{V|U}$ are chosen such that there exists a function $g(U,Y)$ that satisfies $g^n(U^n,Y^n)=\{g(U_{i},Y_{i})\}_{i=1}^n$ and $\mathbb{E}[d(f^n(\widetilde{X}^n,Y^n),g^n(U^n,Y^n))]\leq D+\epsilon_n$, where $\epsilon_n>0$ such that $\epsilon_n\rightarrow 0$ when $n\rightarrow\infty$. 
	Since all sequence tuples $(\widetilde{x}^n,y^n,u^n)$ are in the jointly typical set with high probability, by the typical average lemma \cite[pp. 26]{Elgamalbook}, the distortion constraint (\ref{eq:distortioninnerlossy}) is satisfied. The converse proof follows from the converse proof of Theorem~\ref{theo:innerouter_secrecystorprivregions} by replacing the admissibility step in (\ref{eq:fanoapp}) with the steps
	\begin{align}
		&D+\delta_n\geq \mathbb{E}\Big[d\left(f^n(\widetilde{X}^n,Y^n),\widehat{f^n}(W,Y^n)\right)\Big]\nonumber\\
		&=\frac{1}{n}\mathbb{E}\Big[\sum_{i=1}^nd\left(f_i(\widetilde{X}_i,Y_i),\widehat{f_i}(W,Y^n)\right)\Big]\nonumber\\
		&\overset{(a)}{\geq}\frac{1}{n}\mathbb{E}\Big[\sum_{i=1}^nd\left(f_i(\widetilde{X}_i,Y_i),g_i(W,Y^n,X^{i-1},Z^{i-1},i)\right)\Big]\nonumber\\
		&\overset{(b)}{=}\frac{1}{n}\mathbb{E}\Big[\sum_{i=1}^nd\left(f_i(\widetilde{X}_i,Y_i),g_i(W,Y_{i}^n, X^{i-1},Z^{i-1},i)\right)\Big]\nonumber\\
		&\overset{(c)}{=}\frac{1}{n}\mathbb{E}\Big[\sum_{i=1}^nd\left(f(\widetilde{X}_i,Y_i),g(U_i,i,Y_i)\right)\Big]\label{eq:converseforsinglelossy}
	\end{align}
where $(a)$ follows since there exists a function $g_i(\cdot,\cdot)$ that results in a distortion smaller than or equal to the distortion obtained from $\widehat{f_i}(W,Y^n)$, where the distortion is measured with respect to $f_i(\widetilde{X}_i,Y_i)$ for all $i\!\in\![1\!:\!n]$, because $g_i(\cdot,\cdot)$ has additional inputs, $(b)$ follows from the Markov chain $Y^{i-1}-(X^{i-1},Z^{i-1},W,Y_i,Y_{i+1}^n)-f_i$, and $(c)$ follows from the definition of $U_{i}\triangleq (W,X^{i-1},Y^{n}_{i+1},Z^{i-1})$ given in Section~\ref{subsec:converseofTheorem1}. The cardinality bounds follow by preserving the same probability and conditional entropy values as being preserved in Theorem~\ref{theo:innerouter_secrecystorprivregions} with the addition of preserving the value of $g(U,Y)=g(U,V,Y)$, following from the Markov chain $V-(U,Y)-g(U,Y)$. The region $\mathcal{R}_D$ is convexified by using a time-sharing random variable $Q$. 
\end{IEEEproof}

All rate regions in \cite[Section III]{LifengFCTrans} (and, naturally, all previous rate regions recovered by manipulating the regions in \cite[Section III]{LifengFCTrans}) can be recovered from Theorems~\ref{theo:innerouter_secrecystorprivregions} and \ref{theo:innerouter_secrecystorprivregionslossy} by eliminating the remote source, i.e., assuming $\widetilde{X}^n=X^n$, and by rewriting the secrecy leakage constraint in (\ref{eq:secrecyleakage_cons}) as an equivocation measure rather than a mutual information.

\subsection{Lossless Multi-Function Computation}
We provide inner and outer bounds for the multi-function region $\mathcal{R}_{\text{mf}}$ defined in Section~\ref{subsec:losslesmultifunction} in Theorem~\ref{theo:innerouterlosslesmfc}. The corresponding proof is detailed in Section~\ref{sec:proofofTheoremlosslessmfc}.

\begin{theorem}\label{theo:innerouterlosslesmfc}
	\emph{(Inner Bound):} An achievable multi-function region is the union over all $P_{U_j|\widetilde{X}_j}$ and $P_{V_j|U_j}$ such that $U_j$ is admissible for the function $f_j(\widetilde{X}_j,Y_j)$ for all $j\in[1:J]$ of the rate tuples $(R_{\text{s}}, R_{\text{w},[1:J]},R_{\ell,{\text{Dec}},[1:J]}, R_{\ell,{\text{Eve}}})$ satisfying
	\begin{align}
		&R_{\text{s}}\!\geq\![I(U_{[1:J]};Z_{[1:J]}|V_{[1:J]},Q)\!-\!I(U_{[1:J]};Y_{[1:J]}|V_{[1:J]},Q)]^-\!\nonumber\\
		&\qquad\qquad\quad\!+\!I(U_{[1:J]};\widetilde{X}_{[1:J]}|Z_{[1:J]})\label{eq:secrecyinnerlosslessmfc}\\
		&R_{\text{w},j}\!\geq\! I(U_j;\widetilde{X}_j|Y_j),\qquad\qquad\qquad\qquad\forall j\in[1:J]\label{eq:storageinnerlosslessmfc}\\
		&\sum_{j=1}^{J} R_{\text{w},j} \geq I(U_{[1:J]};\widetilde{X}_{[1:J]}|Y_{[1:J]})\\
		&R_{\ell,\text{Dec},j}\!\geq	\! I(U_j;X|Y_j),\,\qquad\qquad\quad\qquad\forall j\in[1:J]\label{eq:privacyDecinnerlosslessmfc}\\
		&\!R_{\ell,\text{Eve}}\!\geq\![I(U_{[1:J]};Z_{[1:J]}|V_{[1:J]},Q)\!-\!I(U_{[1:J]};Y_{[1:J]}|V_{[1:J]},Q)]^-\!\nonumber\\
		&\qquad\qquad\quad\!+\!I(U_{[1:J]};X|Z_{[1:J]})	\label{eq:privacyEveinnerlosslessmfc}
	\end{align}
		where we have
	\begin{align}
&P_{QV_{[1:J]}U_{[1:J]}\widetilde{X}_{[1:J]}XY_{[1:J]}Z_{[1:J]}}\nonumber\\
&\qquad=P_QP_X\prod_{j=1}^JP_{V_j|U_j}P_{U_j|\widetilde{X}_j}P_{\widetilde{X}_j|X}P_{Y_jZ_j|X}.\label{eq:Theorem3achprobdistribution}
\end{align}

\emph{(Outer Bound):} An outer bound for the multi-function region $\mathcal{R}_{\text{mf}}$ is the union of the rate tuples in (\ref{eq:secrecyinnerlosslessmfc})-(\ref{eq:privacyEveinnerlosslessmfc}) over all $P_{U_{j}|\widetilde{X}_{j}}$ and $P_{V_{j}|U_j}$ such that $U_j$ is admissible for the function $f_j(\widetilde{X}_j,Y_j)$ and $(Q,V_j)-U_j-\widetilde{X}_{j}-X-(Y_j,Z_j)$ form a Markov chain for all $j\in [1:J]$. One can limit the cardinalities to $|\mathcal{Q}|\leq 2$, $|\mathcal{V}_j|\leq |\widetilde{X}_j|+5$, and $|\mathcal{U}_j|\leq (|\widetilde{X}_j|+5)^2$ for all $j\in [1:J]$.

\end{theorem}
\begin{remark}
	The inner and outer bounds differ because the outer bounds define rate regions for the Markov chains $(Q,V_j)-U_j-\widetilde{X}_{j}-X-(Y_j,Z_j)$ for all $j\in [1:J]$, which are larger than the rate regions defined by the inner bounds that satisfy (\ref{eq:Theorem3achprobdistribution}). 
\end{remark}

\subsection{Lossy Multi-Function Computation}
We next give inner and outer bounds for the lossy multi-function region $\mathcal{R}_{\text{mf,D}}$, defined in Section~\ref{subsec:multiplefunctionlossy}, in Theorem~\ref{theo:innerouterlossymfc}.

\begin{theorem}\label{theo:innerouterlossymfc}
	\emph{(Inner Bound):} An achievable lossy multi-function region is the union over all $P_{U_j|\widetilde{X}_j}$ and $P_{V_j|U_j}$ for all $j\in[1:J]$ of the rate tuples $(R_{\text{s}}, R_{\text{w},[1:J]},R_{\ell,{\text{Dec}},[1:J]}, R_{\ell,{\text{Eve}}},D_{[1:J]})$ satisfying 
		\begin{align}
	&R_{\text{s}}\!\geq\![I(U_{[1:J]};Z_{[1:J]}|V_{[1:J]},Q)\!-\!I(U_{[1:J]};Y_{[1:J]}|V_{[1:J]},Q)]^-\!\nonumber\\
	&\qquad\qquad\quad\!+\!I(U_{[1:J]};\widetilde{X}_{[1:J]}|Z_{[1:J]})\label{eq:secrecyinnerlossymfc}\\
	&R_{\text{w},j}\!\geq\! I(U_j;\widetilde{X}_j|Y_j),\qquad\qquad\qquad\qquad\forall j\in[1:J]\label{eq:storageinnerlossymfc}\\
	&\sum_{j=1}^{J} R_{\text{w},j} \geq I(U_{[1:J]};\widetilde{X}_{[1:J]}|Y_{[1:J]})\\
	&R_{\ell,\text{Dec},j}\!\geq	\! I(U_j;X|Y_j),\,\qquad\qquad\quad\qquad\forall j\in[1:J]\label{eq:privacyDecinnerlossymfc}\\
	&\!R_{\ell,\text{Eve}}\!\geq\![I(U_{[1:J]};Z_{[1:J]}|V_{[1:J]},Q)\!-\!I(U_{[1:J]};Y_{[1:J]}|V_{[1:J]},Q)]^-\!\nonumber\\
	&\qquad\qquad\quad\!+\!I(U_{[1:J]};X|Z_{[1:J]})	\label{eq:privacyEveinnerlossymfc}\\
	&D_j\geq \mathbb{E}[d(f_j(\widetilde{X}_j,Y_j),g_j(U_j,Y_j))] \;\;\;\qquad \forall j\in [1:J] \label{eq:distortioninnerlossymfc}
	\end{align}
	for a set of functions $\{g_j(\cdot,\cdot)\}_{j=1}^J$ and where (\ref{eq:Theorem3achprobdistribution}) is satisfied.
	
	\emph{(Outer Bound):} An outer bound for the lossy multi-function region $\mathcal{R}_{\text{mf,D}}$ is the union of the rate tuples in (\ref{eq:secrecyinnerlossymfc})-(\ref{eq:distortioninnerlossymfc}) over all $P_{U_{j}|\widetilde{X}_{j}}$ and $P_{V_{j}|U_j}$ such that $(Q,V_j)-U_j-\widetilde{X}_{j}-X-(Y_j,Z_j)$ form a Markov chain for all $j\in [1:J]$. One can limit the cardinalities to $|\mathcal{Q}|\leq 2$, $|\mathcal{V}_j|\leq |\widetilde{X}_j|~+~6$, and $|\mathcal{U}_j|\leq (|\widetilde{X}_j|+6)^2$ for all $j\in [1:J]$.

\end{theorem}

\begin{IEEEproof}[Proof Sketch]
	The inner bound proof of Theorem~\ref{theo:innerouterlossymfc} follows from the achievability proof of Theorem~\ref{theo:innerouterlosslesmfc}, except that $U_j$'s are not necessarily admissible, and with the addition that $P_{U_j|\widetilde{X}_j}$ and $P_{V_j|U_j}$ are chosen such that there exists a set of functions $\{g_j(U_j,Y_j)\}_{j=1}^J$ that satisfy $g^n_j(U_j^n,Y_j^n)=\{g_j(U_{i,j},Y_{i,j})\}_{i=1}^n$ and $\mathbb{E}[d(f^n_j(\widetilde{X}^n_j,Y^n_j),g^n_j(U^n_j,Y^n_j))]\leq D_j+\epsilon_n$ for all $j\in[1:J]$, where $\epsilon_n>0$ such that $\epsilon_n\rightarrow 0$ when $n\rightarrow\infty$. Since all sequence tuples $(\widetilde{x}_j^n,y_j^n,u_j^n)$ are in the jointly typical set with high probability for all $j\in[1:J]$, by the typical average lemma, the distortion constraints in (\ref{eq:distortioninnerlossymfc}) are satisfied. The outer bound proof of Theorem~\ref{theo:innerouterlossymfc} follows from the converse proof of Theorem~\ref{theo:innerouterlosslesmfc} with the replacement of the admissibility step in (\ref{eq:fanoappmfc}) with the steps given in (\ref{eq:converseforsinglelossy}) for random variables and functions with the indices $j=1,2,\ldots,J$.
\end{IEEEproof}

\section{Information Bottleneck Example}\label{sec:example}
Consider the lossy single-function computation problem and suppose $X-Y-Z$ form a Markov chain. The characterization of the corresponding rate region requires one to maximize a mutual information term upper bounded by another mutual information term that should be minimized simultaneously, i.e., an information bottleneck.

\begin{corollary}\label{corol:XYZMarkovLossySingle}
	The lossy region of Theorem~\ref{theo:innerouter_secrecystorprivregionslossy} when $X-Y-Z$ form a Markov chain is the set of all tuples
	$(R_{\text{s}}, R_{\text{w}},R_{\ell,{\text{Dec}}}, R_{\ell,{\text{Eve}}},D)$ satisfying
	\begin{align}
	&R_{\text{s}}\!\geq  I(U;\widetilde{X}|Y) = I(U;\widetilde{X})-I(U;Y)\label{eq:secrecyinnerlossyexample}\\
	&R_{\text{w}}\!\geq\! I(U;\widetilde{X}|Y)= I(U;\widetilde{X})-I(U;Y)\label{eq:storageinnerlossyexample}\\
	&R_{\ell,\text{Dec}}\!\geq	\! I(U;X|Y)= I(U;X)-I(U;Y)\label{eq:privacyDecinnerlossyexample}\\
	&R_{\ell,\text{Eve}}\!\geq\! I(U;X|Y)= I(U;X)-I(U;Y)	\label{eq:privacyEveinnerlossyexample}\\
	&D\geq \mathbb{E}[d(f(\widetilde{X},Y),g(U,Y))]\label{eq:distortioninnerlossyexample}
	\end{align}
	for some function $g(\cdot,\cdot)$ such that $U-\widetilde{X}-X-Y-Z$ form a Markov chain. One can limit the cardinality to $|\mathcal{U}|\leq |\widetilde{X}|+2$.
\end{corollary}

The proof of Corollary~\ref{corol:XYZMarkovLossySingle} follows by applying steps identical to the proof of \cite[Corollary 3]{LifengFCTrans} to Theorem~\ref{theo:innerouter_secrecystorprivregionslossy}; therefore, we omit it. The boundary points of the rate region defined in Corollary~\ref{corol:XYZMarkovLossySingle} can be obtained by maximizing $I(U;Y)$ and minimizing $I(U;\widetilde{X})$ simultaneously for a fixed $I(U;X)$ for all $P_{U|\widetilde{X}}$ such that $U-\widetilde{X}-X-Y-Z$ form a Markov chain. This problem is an information bottleneck problem \cite{InfoBottleneck,WitsenhausenWynerIB}. If the distortion metric $d(\cdot,\cdot)$ is chosen to be the Hamming distance, we then obtain the optimal function $g^*(u,y)$ for all $(u,y)\!\in\!\mathcal{U}\!\times\!\mathcal{Y}$ as \cite[Eq.~(26)]{LifengFCTrans}
\begin{align}
	g^*(u,y)=\arg\max_{f}P_{F|UY}(f|u,y)
\end{align}
where $f=f(\widetilde{x},y)$ is a realization of the random function output $F$ for any $(\widetilde{x},y)\in\mathcal{\widetilde{X}}\times\mathcal{Y}$. 

Consider a measurement channel $P_{\widetilde{X}|X}$ and source $P_X$ for the encoder $\Enc(\cdot)$ such that the inverse channel $P_{X|\widetilde{X}}$ is a BSC$(p)$ for any $0\leq p\leq 0.5$. Furthermore, suppose the measurement channel $P_{Y|X}$ for the decoder $\Dec(\cdot)$ is a binary input symmetric output channel \cite[p. 21]{gallagerbook}, which can be decomposed into a mixture of binary subchannels as defined in \cite[Section III-B]{bizimCNS} \cite{chayat}. We remark that the rate region defined in Corollary~\ref{corol:XYZMarkovLossySingle} by (\ref{eq:secrecyinnerlossyexample})-(\ref{eq:distortioninnerlossyexample}) does not depend on the random variable $Z$. Therefore, the measurement channel for the eavesdropper does not affect the rate region as long as the measurement channel for the eavesdropper is physically-degraded as compared to the channel for the decoder $\Dec(\cdot)$, i.e., $P_{YZ|X}=P_{Z|Y}P_{Y|X}$. Since $P_{\widetilde{X}XYZ}$ is fixed, the optimal auxiliary random variable $U$ is such that $P_{\widetilde{X}|U}$ is a BSC with crossover probability
\begin{align}
	\frac{H_b^{-1}(H(X|U))-p}{1-2p}\label{eq:optimalUforexample}
\end{align}
which follows from \cite[Theorem 3]{bizimMMMMTIFS}.

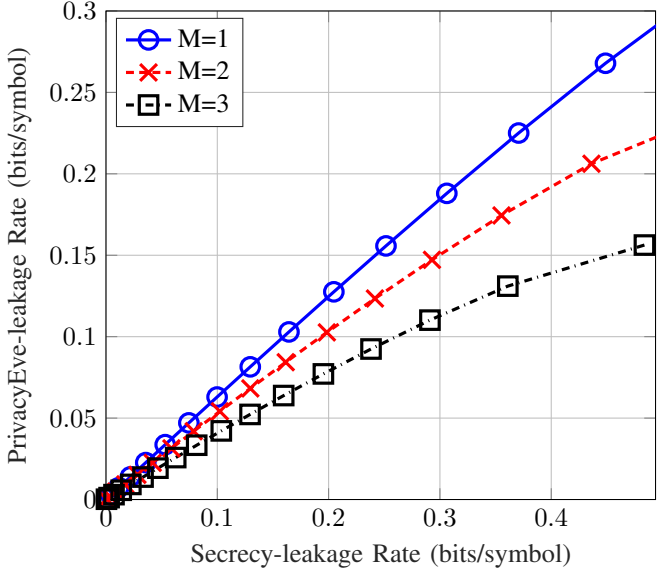
\begin{figure}[t]
	\centering
	\newlength\figureheight
	\newlength\figurewidth
	\setlength\figureheight{9.97cm}
	\setlength\figurewidth{4.8cm}
%
%
\begin{tikzpicture}

\begin{axis}[%
width=7.306cm,
height=6.5cm,
at={(0cm,0cm)},
scale only axis,
xmin=0,
xmax=0.4937,
xlabel style={font=\color{white!15!black}},
xlabel={Secrecy-leakage Rate (bits/symbol)},
ymin=0,
ymax=0.3,
ylabel style={font=\color{white!15!black}},
ylabel={PrivacyEve-leakage Rate (bits/symbol)},
axis background/.style={fill=white},
xmajorgrids,
ymajorgrids,
/pgf/number format/precision=5,
legend style={at={(0.25,0.986)}, legend cell align=left, align=left, draw=white!15!black}
]
\addplot [color=blue, line width=1.2pt, mark size=3.5pt, mark=o, mark options={solid, blue}]
  table[row sep=crcr]{%
0.705636606460175	0.378191687305699\\
0.548007383256479	0.318061301873023\\
0.448749020681236	0.267889679260316\\
0.370725532751793	0.225055506577332\\
0.306188109992181	0.188009196330476\\
0.251536761520942	0.155740459777167\\
0.204705904279172	0.127547780173388\\
0.164371540388158	0.102922475539206\\
0.129625234284984	0.081484218205568\\
0.0998151845046481	0.06294247913476\\
0.0744596762086278	0.0470721906386084\\
0.0531961493793668	0.0336977400973865\\
0.0357495022602763	0.022682116076448\\
0.0219115683289811	0.0139193927681112\\
0.0115274929693276	0.00732947185565891\\
0.00448661585616633	0.00285441728119445\\
0.00071646824693794	0.000455967336613128\\
};
\addlegendentry{M=1}

\addplot [color=red, densely dashed, line width=1.2pt, mark size=4.5pt, mark=x, mark options={solid, red}]
  table[row sep=crcr]{%
0.571580675762127	0.244135756607651\\
0.436141867623629	0.206195786240173\\
0.355309928458748	0.174450587037828\\
0.292857177573966	0.147187151399505\\
0.241625666465081	0.123446752803376\\
0.198423313997053	0.102627012253279\\
0.161478259069309	0.0843201349635246\\
0.129684620950495	0.0682355561015428\\
0.102299649072441	0.0541586329930246\\
0.0787994511159033	0.0419267457460152\\
0.0588020949419155	0.0314146093718961\\
0.0420232302289816	0.0225248209470013\\
0.0282489470696521	0.0151815608858238\\
0.0173184678412245	0.00932629228035453\\
0.00911280633159328	0.00491478521792454\\
0.00354725886690133	0.00191506029192945\\
0.000566502513872913	0.000306001603548101\\
};
\addlegendentry{M=2}

\addplot [color=black, dashdotted, line width=1.2pt, mark size=3.5pt, mark=square, mark options={solid, black}]
  table[row sep=crcr]{%
0.483714723325697	0.15626980417122\\
0.360989943592305	0.131043862208848\\
0.291025008504761	0.110165667083841\\
0.238097879701184	0.0924278535267223\\
0.195315738554472	0.0771368248927674\\
0.159644909447967	0.0638486077041929\\
0.129416948709575	0.0522588246037912\\
0.103597874104751	0.0421488092557991\\
0.0814972299532387	0.0333562138738226\\
0.0626303978438968	0.0257576924740087\\
0.0466455817160128	0.0192580961459934\\
0.0332818416306209	0.0137834323486406\\
0.0223435122199159	0.00927612603608763\\
0.0136839335654941	0.00569175800462413\\
0.00719481504067654	0.0029967939270078\\
0.00279920594756178	0.0011670073725899\\
0.00044691338349867	0.000186412473173858\\
};
\addlegendentry{M=3}

\end{axis}

\begin{axis}[%
width=8.589cm,
height=6.135cm,
at={(-1.117cm,-0.675cm)},
scale only axis,
xmin=0,
xmax=1,
ymin=0,
ymax=1,
axis line style={draw=none},
ticks=none,
axis x line*=bottom,
axis y line*=left,
legend style={legend cell align=left, align=left, draw=white!15!black}
]
\end{axis}
\end{tikzpicture}%
	\caption{Secrecy-leakage rate vs. privacyEve-leakage rate projection of the boundary tuples $(R_{\text{s}}, R_{\text{w}},R_{\ell,{\text{Dec}}}, R_{\ell,{\text{Eve}}},D)$ for $p=0.06$ and for the number of independent BSC measurements at the decoder $M=1,2,3$.} \label{fig:exampleplot}
\end{figure} 

Suppose $P_X\!\sim\!\text{Bern}(0.5)$, $P_{\widetilde{X}|X}\!\sim\!\text{BSC}(p\!=\!0.06)$, and assume that the measurement channel $P_{Y|X}$ consists of $M> 1$ independent BSCs each with crossover probability $0.15$, which satisfies the assumptions listed above. Using auxiliary random variables satisfying (\ref{eq:optimalUforexample}), we depict the projections of $(R_{\text{s}}, R_{\text{w}},R_{\ell,{\text{Dec}}}, R_{\ell,{\text{Eve}}},D)$ boundary tuples onto the $(R_{\text{s}}, R_{\ell,{\text{Eve}}}) $ plane in Fig.~\ref{fig:exampleplot} for $M=1,2,3$ independent BSC measurements by the decoder $\Dec(\cdot)$.

Fig.~\ref{fig:exampleplot} suggests that given a boundary point achieved by a crossover probability calculated as in (\ref{eq:optimalUforexample}), any larger secrecy-leakage rate and any larger privacyEve-leakage rate are also achievable. Conversely, given such an achievable boundary point, no smaller secrecy-leakage rate and no smaller privacyEve-leakage rate is achievable. Furthermore, increasing the number $M$ of measurements at the decoder significantly decreases the corresponding boundary point such that, e.g., when $M=3$ measurements are used as compared to $M=1$, the maximum secrecy-leakage rate decreases by approximately $31.45\%$ and simultaneously the maximum privacy-leakage rate to the eavesdropper decreases by approximately $58.68\%$. These gains can be seen as multiplexing gains, in analogy to multiple antenna systems for wireless communications.

\section{Proof of Theorem~\ref{theo:innerouter_secrecystorprivregions}}\label{sec:proofofTheorem1}\label{sec:proofTheorem1}
\subsection{Achievability Proof of Theorem~\ref{theo:innerouter_secrecystorprivregions}}\label{subsec:achproofofTheorem1}
\begin{IEEEproof}[Proof Sketch]
	We use the output statistics of random binning (OSRB) method, proposed in \cite{AminOSRB} (see also \cite{RenesRenner}) for strong secrecy by following steps in \cite[Section~1.6]{BlochLectureNotes2018}. This approach simplifies the analysis compared to previous proofs in the literature.
	
	Fix $P_{U|\widetilde{X}}$ and $P_{V|U}$ such that $U$ is admissible and let $(V^n,U^n,\widetilde{X}^n,X^n,Y^n,Z^n)$ be i.i.d. according to $P_{VU\widetilde{X}XYZ}=P_{V|U}P_{U|\widetilde{X}}P_{\widetilde{X}|X}P_XP_{YZ|X}$. We remark that since all $n$-letter random variables are i.i.d., $U^n$ is also admissible.

Assign two random bin indices $(F_{\text{v}},W_{\text{v}})$	to each $v^n$. Assume $F_{\text{v}}\in[1:2^{n\widetilde{R}_{\text{v}}}]$ and $W_{\text{v}}\in[1:2^{nR_{\text{v}}}]$. Similarly, assign two indices $(F_{\text{u}},W_{\text{u}})$ to each $u^n$, where $F_{\text{u}}\in[1:2^{n\widetilde{R}_{\text{u}}}]$ and $W_{\text{u}}\in[1:2^{nR_{\text{u}}}]$. The public message is $W=(W_{\text{v}}, W_{\text{u}})$ and the indices $F=(F_{\text{v}}, F_{\text{u}})$ represent the public choice of encoder-decoder pairs.

Using a Slepian-Wolf (SW) \cite{SW} decoder one can reliably estimate $V^n$ from $(F_{\text{v}},W_{\text{v}}, Y^n)$, such that the expected value of the error probability taken over the random bin assignments vanishes when $n\rightarrow\infty$, if we have \cite[Lemma~1]{OSRBAmin}
\begin{align}
\widetilde{R}_{\text{v}} + R_{\text{v}}> H(V|Y).\label{eq:Vnrecons}
\end{align}	
Similarly, one can reliably estimate $U^n$ from $(F_{\text{u}},W_{\text{u}}, Y^n, V^n)$ by using a SW decoder if we have
\begin{align}
\widetilde{R}_{\text{u}} + R_{\text{u}}> H(U|V,Y).\label{eq:Unrecons}
\end{align}
Thus, the reliability constraint in (\ref{eq:reliability_cons}) is satisfied if (\ref{eq:Vnrecons}) and (\ref{eq:Unrecons}) are satisfied.

The public index $F_{\text{v}}$ is almost independent of $\widetilde{X}^n$, so it is almost independent of $(\widetilde{X}^n,X^n,Y^n,Z^n)$, if we have \cite[Theorem 1]{OSRBAmin}
\begin{align}
\widetilde{R}_{\text{v}}<H(V|\widetilde{X})\label{eq:independenceofFv}
\end{align}
since it results in  the expected value, which is taken over the random bin assignments, of the variational distance between the joint probability distributions $\text{Unif}[1\!\!:\!|\mathcal{F}_{\text{v}}|]\cdot\text{Unif}[1 \!\! :\!\!|\mathcal{\widetilde{X}}|^n]$ and $P_{F_{\text{v}}\widetilde{X}^n}$ to vanish when $n\rightarrow\infty$. Furthermore, the public index $F_{\text{u}}$ is almost independent of $(V^n,\widetilde{X}^n)$, so it is almost independent of $(V^n,\widetilde{X}^n,X^n,Y^n,Z^n)$, if we have 
\begin{align}
\widetilde{R}_{\text{u}}<H(U|V,\widetilde{X}).\label{eq:independenceofFu}
\end{align} 

To satisfy the constraints (\ref{eq:Vnrecons})-(\ref{eq:independenceofFu}), we fix the rates to
\begin{align}
&\widetilde{R}_{\text{v}} = H(V|\widetilde{X})-\epsilon\label{eq:R_vtildechosen}\\
&R_{\text{v}} = I(V;\widetilde{X})-I(V;Y)+2\epsilon\label{eq:R_vchosen}\\
&\widetilde{R}_{\text{u}} = H(U|V,\widetilde{X})-\epsilon\label{eq:R_utildechosen}\\
&R_{\text{u}} = I(U;\widetilde{X}|V)-I(U;Y|V)+2\epsilon\label{eq:R_uchosen}
\end{align}
for any $\epsilon>0$. 

\textbf{Storage (Public Message) Rate}: (\ref{eq:R_vtildechosen})-(\ref{eq:R_uchosen}) result in a storage (public message) rate $R_{\text{w}}$ of
\begin{align}
&R_{\text{w}} = R_{\text{v}} + R_{\text{u}}=  I(V, U;\widetilde{X})-I(V,U;Y)+4\epsilon\nonumber\\
&\; \overset{(a)}{=}   I(U;\widetilde{X}|Y)+4\epsilon\label{eq:R_wchosen}
\end{align}
where $(a)$ follows because $V-U-\widetilde{X}-Y$ form a Markov chain.

\textbf{Privacy Leakage to the Decoder}: We have
\begin{align}
&I(X^n;W,F|Y^n) = I(X^n;W|F,Y^n)+I(X^n;F|Y^n)\nonumber\\
&\; \overset{(a)}{\leq} H(X^n|Y^n) - H(X^n|W,F,V^n, U^n,Y^n)+2\epsilon_n\nonumber\\
&\;\overset{(b)}{=}H(X^n|Y^n) - H(X^n|U^n,Y^n)+2\epsilon_n\nonumber\\
&\; \overset{(c)}{=}nI(U;X|Y)+2\epsilon_n\label{eq:ach_privtoDec}
\end{align}
where $(a)$ follows for some $\epsilon_n>0$ with $\epsilon_n\rightarrow 0$ when $n\rightarrow\infty$ because
\begin{align}
&I(X^n;F|Y^n) = I(X^n;F_\text{v}|Y^n)+I(X^n;F_\text{u}|F_\text{v},Y^n)\nonumber\\
&\;\leq 2\epsilon_n\label{eq:XnYnandFareind}
\end{align}
since by (\ref{eq:independenceofFv}) $F_v$ is almost independent of $(\widetilde{X}^n,X^n,Y^n,Z^n)$ and by (\ref{eq:independenceofFu}) $F_u$ is almost independent of $(V^n,\widetilde{X}^n,X^n,Y^n,Z^n)$ and because $V^n$ determines $F_v$, $(b)$ follows because $V^n$ determines $(F_v,W_v)$, $U^n$ determines $(F_u,W_u)$, and $V^n-U^n-(X^n,Y^n)$ form a Markov chain, and $(c)$ follows because $(X^n,U^n,Y^n)$ are i.i.d.

\textbf{Privacy Leakage to the Eavesdropper}: We obtain
\begin{align}
&I(X^n;W,F|Z^n)\overset{(a)}{=}H(W,F|Z^n)-H(W,F|X^n)\nonumber\\
&\; \overset{(b)}{=} H(W,F|Z^n)-H(W_u,F_u,V^n|X^n)\nonumber\\
&\qquad + H(V^n|W_v,F_v,W_u,F_u,X^n)\nonumber
\end{align}
\begin{align}
&\; \overset{(c)}{\leq} H(W,F|Z^n)-H(W_u,F_u,V^n|X^n)+n\epsilon^{\prime}_n\nonumber\\
&\;\overset{(d)}{=}H(W,F|Z^n)-H(U^n,V^n|X^n)\nonumber\\
&\qquad+H(U^n|W_u,F_u,V^n,X^n)+n\epsilon^{\prime}_n\nonumber\\
&\;\overset{(e)}{\leq}H(W,F|Z^n)-H(U^n,V^n|X^n)+2n\epsilon^{\prime}_n\nonumber\\
&\;\overset{(f)}{=}H(W,F|Z^n)-nH(U,V|X)+2n\epsilon^{\prime}_n\label{eq:ach_privtoEvefirststep}
\end{align}
where $(a)$ follows because $(W,F)-X^n-Z^n$ form a Markov chain, $(b)$ follows since $V^n$ determines $(F_v,W_v)$, $(c)$ follows for some $\epsilon^{\prime}_n>0$ such that $\epsilon^{\prime}_n\rightarrow0$ when $n\rightarrow\infty$ because $(F_v,W_v,X^n)$ can reliably recover $V^n$ due to the Markov chain $V^n-X^n-Y^n$ and by (\ref{eq:Vnrecons}), $(d)$ follows because $U^n$ determines $(F_u,W_u)$, $(e)$ follows by (\ref{eq:Unrecons})  because $(W_u,F_u,V^n,X^n)$ can reliably recover $U^n$ due to the  inequality $H(U|V,Y)\geq H(U|V,X)$ that follows from
\begin{align}
& H(U|V,Y) - H(U|V,X)= I(U;V,X)-I(U;V,Y)\nonumber\\
&\;\geq I(U;V,X)-I(U;V,Y,X)= 0\label{eq:proofthatHUVYisgreaterthanHUVX}
\end{align}
since $(V,U)-X-Y$ form a Markov chain, and $(f)$ follows because $(U^n,V^n,X^n)$ are i.i.d. 

We need to analyze six different decodability cases to consider whether $(F_v,W_v,Z^n)$ can recover $V^n$ and whether $(F_u,W_u,V^n,Z^n)$ or $(F_u,W_u,Z^n)$ can recover $U^n$.

\emph{Case 1}: Assume
\begin{align}
0\leq&\;R_\text{v}+\widetilde{R}_\text{v} < H(V|Z) \label{eq:case1_constv},\\
0\leq&\;R_\text{u}+\widetilde{R}_\text{u} < H(U|V,Z)\label{eq:case1_constu}
\end{align}
so that $(F_v,W_v)$ are almost independent of $Z^n$ and are also almost mutually independent, and $(F_u,W_u)$ are almost independent of $(V^n,Z^n)$ and are also almost mutually independent. Using (\ref{eq:ach_privtoEvefirststep}), we obtain
\begin{align}
&I(X^n;W,F|Z^n)\nonumber\\
&\leq \!H(W_v)\!+\!H(F_v)\!+\!H(W_u)\!+\!H(F_u)\!-\!nH(U,V|X)\!+\!2n\epsilon^{\prime}_n\nonumber\\
&\leq n(R_\text{v}+\widetilde{R}_\text{v}+R_\text{u}+\widetilde{R}_\text{u})\!-\!nH(U,V|X)\!+\!2n\epsilon^{\prime}_n\nonumber\\
& \overset{(a)}{=} n(I(U,V;X)-I(U,V;Y)+2\epsilon+2\epsilon^{\prime}_n)\nonumber\\
& \overset{(b)}{=} n(I(U;X)-I(U;Y|V)-I(V;Y)+2\epsilon+2\epsilon^{\prime}_n)\nonumber\\
& \overset{(c)}{\leq} n(I(U;X)-I(U;Y|V)-I(V;Z)+\epsilon+2\epsilon^{\prime}_n)\nonumber\\
& \overset{(d)}{=} \!n(I(U;X)\!-\![I(U;Y|V)\!-\!I(U;Z|V)]\!-\!I(U;Z)\!+\!\epsilon\!+\!2\epsilon^{\prime}_n)\nonumber\\
& \overset{(e)}{=}\! \!n(I(U;X|Z)\!+\![I(U;Z|V)\!-\!I(U;Y|V)\!+\!\epsilon]^-\!+\!2\epsilon^{\prime}_n)\label{eq:ach_privtoEveCase1}
\end{align}
where $(a)$ follows by (\ref{eq:R_vtildechosen})-(\ref{eq:R_uchosen}) and $(b)$ follows from the Markov chain $V-U-X$, $(c)$ follows by (\ref{eq:R_vtildechosen}), (\ref{eq:R_vchosen}), and (\ref{eq:case1_constv}) such that equality is achieved when $n\rightarrow\infty$, $(d)$ follows from the Markov chain $V-U-Z$, and $(e)$ follows from the Markov chain $U-X-Z$.

\emph{Case 2}: Assume
\begin{align}
0\leq&\;R_\text{v}+\widetilde{R}_\text{v} < H(V|Z) \label{eq:case2_constv},\\
H(U|V,Z)< &\; R_\text{u}+\widetilde{R}_\text{u} < H(U|Z)\label{eq:case2_constu}
\end{align}
so that $(F_v,W_v)$ are almost independent of $Z^n$ and are also almost mutually independent, and $(F_u,W_u)$ are almost independent of $Z^n$ and are also almost mutually independent; however, $(F_u,W_u,V^n,Z^n)$ can reliably recover $U^n$. Using (\ref{eq:ach_privtoEvefirststep}), we have
\begin{align}
&I(X^n;W,F|Z^n)\nonumber\\
&\!\overset{(a)}{\leq} H(U^n,V^n|Z^n)\!-\!nH(U,V|X)\!+\!2n\epsilon^{\prime}_n\nonumber\\
&\!\overset{(b)}{=}\!n(I(U;X|Z)\!+\![I(U;Z|V)\!-\!I(U;Y|V)\!+\!\epsilon]^-\!+\!2\epsilon_n^{\prime})\label{eq:ach_privtoEveCase2}
\end{align}
where $(a)$ follows because $V^n$ determines $(F_v,W_v)$ and $U^n$ determines $(F_u,W_u)$, and $(b)$ follows because $(V^n,U^n,Z^n)$ are i.i.d., from the Markov chain $V-U-X-Z$, and by (\ref{eq:R_utildechosen}), (\ref{eq:R_uchosen}), and (\ref{eq:case2_constu}).

\emph{Case 3}: Assume
\begin{align}
0\leq&\;R_\text{v}+\widetilde{R}_\text{v} < H(V|Z) \label{eq:case3_constv},\\
H(U|Z)< &\; R_\text{u}+\widetilde{R}_\text{u} \label{eq:case3_constu}
\end{align}
so that $(F_v,W_v)$ are almost independent of $Z^n$ and are also almost mutually independent, and $(F_u,W_u,Z^n)$ can reliably recover $U^n$. Using (\ref{eq:ach_privtoEvefirststep}), we obtain
\begin{align}
&I(X^n;W,F|Z^n)\nonumber\\
&\!\overset{(a)}{\leq} H(U^n|Z^n)\!+\!H(W_v,F_v|U^n,Z^n)\!-\!nH(U,V|X)\!+\!2n\epsilon^{\prime}_n\nonumber\\
&\!\overset{(b)}{\leq} H(U^n|Z^n)\!+\!H(V^n|U^n,Z^n)\!-\!nH(U,V|X)\!+\!2n\epsilon^{\prime}_n\nonumber\\
&\!\overset{(c)}{=} n(I(U;X|Z)\!+\!2\epsilon^{\prime}_n)\nonumber\\
&\!\overset{(d)}{=}\! n(I(U;X|Z)\!+\![I(U;Z|V)\!-\!I(U;Y|V)\!+\!\epsilon]^-\!+\!\!2\epsilon^{\prime}_n)\label{eq:ach_privtoEveCase3}
\end{align}
where $(a)$ follows because $U^n$ determines $(F_u,W_u)$, $(b)$ follows since $V^n$ determines $(F_v,W_v)$, $(c)$ follows from the Markov chain $V-U-X-Z$ and because $(V^n,U^n,X^n,Z^n)$ are i.i.d., and $(d)$ follows by (\ref{eq:R_utildechosen}), (\ref{eq:R_uchosen}), and (\ref{eq:case3_constu}).

\emph{Case 4}: Assume
\begin{align}
H(V|Z) <&\;R_\text{v}+\widetilde{R}_\text{v} \label{eq:case4_constv},\\
0\leq&\;R_\text{u}+\widetilde{R}_\text{u} < H(U|V,Z)\label{eq:case4_constu}
\end{align}
so that $(F_v,W_v,Z^n)$ can reliably recover $V^n$, and $(F_u,W_u)$ are almost independent of $(V^n,Z^n)$ and are also almost mutually independent. Using (\ref{eq:ach_privtoEvefirststep}), we have
\begin{align}
&I(X^n;W,F|Z^n)\nonumber\\
& \overset{(a)}{\leq}H(V^n|Z^n)+H(W_u,F_u|W_v,F_v,Z^n)\nonumber\\
&\qquad-nH(U,V|X)+2n\epsilon^{\prime}_n\nonumber\\
&\leq H(V^n|Z^n)+H(W_u)+H(F_u)-nH(U,V|X)+2n\epsilon^{\prime}_n\nonumber\\
&\leq n(H(V|Z)+R_{\text{u}}+\widetilde{R}_{\text{u}}-H(U,V|X)+2\epsilon^{\prime}_n)\nonumber\\
&\overset{(b)}{=}n(H(V|Z)+H(U|V,Y)+\epsilon-H(U,V|X)+2\epsilon^{\prime}_n)\nonumber\\
&=n(I(U;X|V)\!-\!I(U;Y|V)\!+\! I(V;X)\!-\!I(V;Z)\!+\!2\epsilon^{\prime}_n\!+\!\epsilon)\nonumber\\
&\overset{(c)}{=}n(I(U;X)-I(U;Y|V)-I(V;Z)+2\epsilon^{\prime}_n+\epsilon\nonumber\\
& \overset{(d)}{=} \!n(I(U;X)\!-\![I(U;Y|V)\!-\!I(U;Z|V)]\!-\!I(U;Z)\!+\!\epsilon\!+\!2\epsilon^{\prime}_n)\nonumber\\
& \overset{(e)}{=}\! \!n(I(U;X|Z)\!+\![I(U;Z|V)\!-\!I(U;Y|V)\!+\!\epsilon]^-\!+\!2\epsilon^{\prime}_n)\label{eq:ach_privtoEveCase4}
\end{align}
where $(a)$ follows because $V^n$ determines $(F_v,W_v)$, $(b)$ follows because $(V^n,Z^n)$ are i.i.d. and by (\ref{eq:R_utildechosen}) and (\ref{eq:R_uchosen}), $(c)$ follows from the Markov chain $V-U-X$, $(d)$ follows from the Markov chain $V-U-Z$, and $(e)$ follows from the Markov chain $U-X-Z$.

\emph{Case 5}: Assume
\begin{align}
H(V|Z) <&\;R_\text{v}+\widetilde{R}_\text{v}  \label{eq:case5_constv},\\
H(U|V,Z)< &\; R_\text{u}+\widetilde{R}_\text{u} < H(U|Z)\label{eq:case5_constu}
\end{align}
so that $(F_v,W_v,Z^n)$ can reliably recover $V^n$,  and $(F_u,W_u)$ are almost independent of $Z^n$ and are also almost mutually independent; however, $(F_u,W_u,V^n,Z^n)$ can reliably recover $U^n$. Using (\ref{eq:ach_privtoEvefirststep}), we have
\begin{align}
&I(X^n;W,F|Z^n)\nonumber\\
&\overset{(a)}{\leq} H(V^n|Z^n)+H(W_u,F_u|W_v,F_v,Z^n)\nonumber\\
&\qquad -nH(U,V|X)+2n\epsilon^{\prime}_n\nonumber\\
&\overset{(b)}{\leq}H(V^n|Z^n)+H(W_u,F_u|V^n,Z^n)+H(V^n|W_v,F_v,Z^n)\nonumber\\
&\qquad -nH(U,V|X)+2n\epsilon^{\prime}_n\nonumber\\
&\overset{(c)}{\leq}H(V^n|Z^n)+H(U^n|V^n,Z^n) -nH(U,V|X)+3n\epsilon^{\prime}_n\nonumber\\
&\overset{(d)}{=} n(I(U;X|Z)\!+\!3\epsilon^{\prime}_n)\nonumber\\
&\overset{(e)}{=}\! n(I(U;X|Z)\!+\![I(U;Z|V)\!-\!I(U;Y|V)\!+\!\epsilon]^-\!+\!\!3\epsilon^{\prime}_n)\label{eq:ach_privtoEveCase5}
\end{align}
where $(a)$ and $(b)$ follow because $V^n$ determines $(F_v,W_v)$, $(c)$ follows because $U^n$ determines $(F_u,W_u)$ and by (\ref{eq:case5_constv}), $(d)$ follows because $(V^n,Z^n,U^n)$ are i.i.d. and from the Markov chain $V-U-X-Z$, and $(e)$ follows by (\ref{eq:R_utildechosen}), (\ref{eq:R_uchosen}), and (\ref{eq:case5_constu}).

\emph{Case 6}: Assume
\begin{align}
H(V|Z) <&\; R_\text{v}+\widetilde{R}_\text{v} \label{eq:case6_constv},\\
H(U|Z)< &\; R_\text{u}+\widetilde{R}_\text{u} \label{eq:case6_constu}
\end{align}
so that $(F_v,W_v,Z^n)$ can reliably recover $V^n$, and $(F_u,W_u,Z^n)$ can reliably recover $U^n$. Using (\ref{eq:ach_privtoEvefirststep}), we obtain
\begin{align}
&I(X^n;W,F|Z^n)\overset{(a)}{\leq} H(V^n,U^n|Z^n)-nH(U,V|X)+2n\epsilon^{\prime}_n\nonumber\\
&\overset{(b)}{=}n(I(U;X|Z)+2\epsilon^{\prime}_n)\nonumber\\
&\overset{(c)}{=}\! n(I(U;X|Z)\!+\![I(U;Z|V)\!-\!I(U;Y|V)\!+\!\epsilon]^-\!+\!\!2\epsilon^{\prime}_n)\label{eq:ach_privtoEveCase6}
\end{align}
where $(a)$ follows because $U^n$ determines $(F_u,W_u)$ and $V^n$ determines $(F_v,W_v)$, $(b)$ follows because $(V^n,U^n,Z^n)$ are i.i.d. and from the Markov chain $V-U-X-Z$, and $(c)$ follows by (\ref{eq:R_utildechosen}), (\ref{eq:R_uchosen}), and (\ref{eq:case6_constu}).

\textbf{Secrecy Leakage (to the Eavesdropper)}: Consider the secrecy leakage. We have
\begin{align}
&I(\widetilde{X}^n,Y^n;W,F|Z^n)\overset{(a)}{=}H(W,F|Z^n)-H(W,F|\widetilde{X}^n)\nonumber\\
&\; \overset{(b)}{\leq} H(W,F|Z^n)-H(W_u,F_u,V^n|\widetilde{X}^n)+n\epsilon^{\prime}_n\nonumber\\
&\;\overset{(c)}{\leq}H(W,F|Z^n)-nH(U,V|\widetilde{X})+2n\epsilon^{\prime}_n\label{eq:ach_secrecyfirststep}
\end{align}
where $(a)$ follows from the Markov chain $(W,F)-\widetilde{X}^n-(Y^n,Z^n)$, $(b)$ follows since $(W_v,F_v,\widetilde{X}^n)$ can reliably recover $V^n$ due to the Markov chain $V^n-\widetilde{X}^n-Y^n$ and (\ref{eq:Vnrecons}), and $(c)$ follows by (\ref{eq:Unrecons}) since $(W_u,F_u,V^n,\widetilde{X}^n)$ can reliably recover $U^n$ due to the inequality $H(U|V,Y)\geq H(U|V,\widetilde{X})$ that can be proved similarly as in (\ref{eq:proofthatHUVYisgreaterthanHUVX}), and because $(U^n,V^n,\widetilde{X}^n)$ are i.i.d.

Similar to the analysis of the privacy leakage to the eavesdropper, we need to analyze the same six decodability cases to consider whether $(F_v,W_v,Z^n)$ can recover $V^n$ and whether $(F_u,W_u,V^n,Z^n)$ or $(F_u,W_u,Z^n)$ can recover $U^n$. One can show that all steps applied in Cases 1-6 for the privacy leakage to the eavesdropper follow also for the Cases 1-6 for the secrecy leakage by replacing $X$ with $\widetilde{X}$. We; therefore, list the results for Cases 1-6 as follows.

\textit{Case 1}: We obtain for (\ref{eq:case1_constv}) and (\ref{eq:case1_constu}) that
\begin{align}
&I(\widetilde{X}^n,Y^n;W,F|Z^n)\nonumber\\
&\!\leq\! n(I(U;\widetilde{X}|Z)\!+\![I(U;Z|V)\!-\!I(U;Y|V)\!+\!\epsilon]^-\!\!+\!2\epsilon^{\prime}_n)\label{eq:ach_secrecyleakCase1}.
\end{align}

\textit{Case 2}: We obtain for (\ref{eq:case2_constv}) and (\ref{eq:case2_constu}) that
\begin{align}
&I(\widetilde{X}^n,Y^n;W,F|Z^n)\nonumber\\
&\!\leq\! n(I(U;\widetilde{X}|Z)\!+\![I(U;Z|V)\!-\!I(U;Y|V)\!+\!\epsilon]^-\!\!+\!2\epsilon^{\prime}_n)\label{eq:ach_secrecyleakCase2}.
\end{align}

\textit{Case 3}: We obtain for (\ref{eq:case3_constv}) and (\ref{eq:case3_constu}) that
\begin{align}
&I(\widetilde{X}^n,Y^n;W,F|Z^n)\nonumber\\
&\!\leq\! n(I(U;\widetilde{X}|Z)\!+\![I(U;Z|V)\!-\!I(U;Y|V)\!+\!\epsilon]^-\!+\!\!2\epsilon^{\prime}_n)\label{eq:ach_secrecyleakCase3}.
\end{align}

\textit{Case 4}:  We obtain for (\ref{eq:case4_constv}) and (\ref{eq:case4_constu}) that
\begin{align}
&I(\widetilde{X}^n,Y^n;W,F|Z^n)\nonumber\\
&\!\leq\! n(I(U;\widetilde{X}|Z)\!+\![I(U;Z|V)\!-\!I(U;Y|V)\!+\!\epsilon]^-\!+\!\!2\epsilon^{\prime}_n)\label{eq:ach_secrecyleakCase4}.
\end{align}

\textit{Case 5}: We obtain for (\ref{eq:case5_constv}) and (\ref{eq:case5_constu}) that
\begin{align}
&I(\widetilde{X}^n,Y^n;W,F|Z^n)\nonumber\\
&\!\leq\!\!n(I(U;\widetilde{X}|Z)\!+\![I(U;Z|V)\!-\!I(U;Y|V)\!+\!\epsilon]^-\!+\!\!3\epsilon^{\prime}_n) \label{eq:ach_secrecyleakCase5}.
\end{align}

\textit{Case 6}: We obtain for (\ref{eq:case6_constv}) and (\ref{eq:case6_constu}) that
\begin{align}
&I(\widetilde{X}^n,Y^n;W,F|Z^n)\nonumber\\
&\!\leq\! \!n(I(U;\widetilde{X}|Z)\!+\![I(U;Z|V)\!-\!I(U;Y|V)\!+\!\epsilon]^-\!+\!\!2\epsilon^{\prime}_n)\label{eq:ach_secrecyleakCase6}.
\end{align}

Now assume that the public indices $F$ are generated uniformly at random. The encoder $\Enc(\cdot)$ generates $(V^n,U^n)$ according to $P_{V^nU^n|\widetilde{X}^nF_{\text{v}}F_{\text{u}}}$ obtained from the binning scheme above to compute the bins $W_{\text{v}}$ from $V^n$ and $W_{\text{u}}$ from $U^n$, respectively. This procedure induces a joint probability distribution that is almost equal to $P_{VU\widetilde{X}XYZ}$ fixed above \cite[Section 1.6]{BlochLectureNotes2018}. We remark that the privacy and secrecy leakage metrics considered above are expectations over all possible realizations $F=f$. Thus, applying the selection lemma \cite[Lemma~2.2]{Blochbook} to each decodability case separately, these results prove the achievability for Theorem~\ref{theo:innerouter_secrecystorprivregions} by choosing an $\epsilon>0$ such that $\epsilon\rightarrow 0$ when $n\rightarrow\infty$.
\end{IEEEproof}

\subsection{Converse Proof of Theorem~\ref{theo:innerouter_secrecystorprivregions}}\label{subsec:converseofTheorem1}

\begin{IEEEproof}[Proof Sketch]
	Suppose for some $\delta_n\!>\!0$ and $n\geq 1$, there exists a pair of encoders and decoders such that (\ref{eq:reliability_cons})-(\ref{eq:privEve_cons}) are satisfied for some tuple $(R_{\text{s}}, R_{\text{w}},R_{\ell,{\text{Dec}}}, R_{\ell,{\text{Eve}}})$.  
	
	Let $V_{i}\triangleq (W,Y^{n}_{i+1},Z^{i-1})$ and $U_{i}\triangleq (W,X^{i-1},Y^{n}_{i+1},Z^{i-1})$, which satisfy the Markov chain $V_i-U_i-\widetilde{X}_i-X_i-(Y_i,Z_i)$ for all $i\in[1:n]$ by definition of the source statistics. 
	
	\textbf{Admissibility of $\mathbf{U}$}: Define $\epsilon_n\!=\!\delta_n |\mathcal{\widetilde{X}}||\mathcal{Y}| \!+\!H_b(\delta_n)/n$, where $H_b(\delta) = -\delta\log \delta - (1-\delta)\log(1-\delta)$ is the binary entropy function, so that $\epsilon_n\!\rightarrow\!0$ if $\delta_n\!\rightarrow\!0$. Using (\ref{eq:reliability_cons}) and Fano's inequality, we obtain
	\begin{align}
	&n\epsilon_n\geq H(f^n|\widehat{f^n})\overset{(a)}{=}H(f^n|\widebar{f}^n)\overset{(b)}{=}\sum_{i=1}^nH(f_i|\widebar{f}_i)\nonumber\\
	&\;\geq\sum_{i=1}^nH(f_i|\widebar{f}^n)\overset{(c)}{\geq}\sum_{i=1}^n H(f_i|W,Y^n)\nonumber\\
	&\;\geq \sum_{i=1}^nH(f_i|W,Y^n,X^{i-1}, Z^{i-1})\nonumber\\
	&\;\overset{(d)}{=}\sum_{i=1}^nH(f_i|W,Y^n_{i+1},X^{i-1}, Z^{i-1},Y_i)\nonumber\\
	&\;\overset{(e)}{=}\sum_{i=1}^nH(f_i|U_i,Y_i)\label{eq:fanoapp} 
	\end{align}
	where $(a)$ follows from \cite[Lemma 2]{CorrelatedPaperLong} so that when $n\rightarrow\infty$, there exists an i.i.d. random variable $\widebar{f}^n$ such that $H(f^n|\widehat{f^n})=H(f^n|\widebar{f}^n)$ and $\widehat{f^n}-\widebar{f}^n-(W,Y^n)$ form a Markov chain, $(b)$ follows because $(f^n,\widebar{f}^n)$ are i.i.d., $(c)$ follows from the Markov chain $f^n-(W,Y^n)-\widebar{f}^n$ and permits randomized decoding, $(d)$ follows from the Markov chain  for all $i\in[1:n]$
	\begin{align}
		Y^{i-1}-(X^{i-1},Z^{i-1},W,Y_i,Y_{i+1}^n)-f_i \label{eq:conv_Markov2}
	\end{align}
	and $(e)$ follows from the definition of $U_i$.
	
	\textbf{Storage (Public Message) Rate}: We have
	\begin{align}
	&n(R_\text{w}+\delta_n) \overset{(a)}{\geq} \log|\mathcal{W}| \geq H(W|Y^n)-H(W|\widetilde{X}^n,Y^n)\nonumber\\
	& =I(\widetilde{X}^n;W|Y^n) = H(\widetilde{X}^n|Y^n)-H(\widetilde{X}^n|W,Y^n)\nonumber\\
	& =H(\widetilde{X}^n|Y^n)-\sum_{i=1}^nH(\widetilde{X}_i|\widetilde{X}^{i-1},W,Y^n)\nonumber\\
	&\overset{(b)}{=}H(\widetilde{X}^n|Y^n)-\sum_{i=1}^nH(\widetilde{X}_i|\widetilde{X}^{i-1},W,Y_{i+1}^n,Y_i)\nonumber\\
	&\overset{(c)}{\geq}H(\widetilde{X}^n|Y^n)-\sum_{i=1}^nH(\widetilde{X}_i|X^{i-1},Z^{i-1},W,Y_{i+1}^n,Y_i)\nonumber\\
	&\overset{(d)}{=}nH(\widetilde{X}|Y)-\sum_{i=1}^nH(\widetilde{X}|U_i,Y_i) =\sum_{i=1}^n I(U_i;\widetilde{X}_i|Y_i)\label{eq:storagerateconv}
	\end{align}
	where $(a)$ follows by (\ref{eq:storage_cons}), $(b)$ follows from the Markov chain for all $i\in[1:n]$
	\begin{align}
		Y^{i-1}-(\widetilde{X}^{i-1},W,Y_{i+1}^n,Y_i)-\widetilde{X}_i\label{eq:markovyiminus1xtildei}
	\end{align}
	$(c)$ follows from the data processing inequality applied to the Markov chain for all $i\in[1:n]$
	\begin{align}
		&(X^{i-1},Z^{i-1})-(\widetilde{X}^{i-1}, W,Y_{i+1}^n,Y_i)-\widetilde{X}_i\label{eq:Markovconversextildeandximinus1}
	\end{align}
	and $(d)$ follows from the definition of $U_i$.

	\textbf{Privacy Leakage to the Decoder}: We obtain
	\begin{align}
	&n(R_{\ell,\text{Dec}}+\delta_n) \overset{(a)}{\geq}H(W|Y^n)-H(W|X^n)\nonumber\\
	&\; =\sum_{i=1}^n \Big[I(W;X_i|X^{i-1}) - I(W;Y_i|Y_{i+1}^n)\Big]\nonumber\\
	&\; \overset{(b)}{=}\sum_{i=1}^n \Big[I(W;X_i|X^{i-1},Y_{i+1}^n) - I(W;Y_i|Y_{i+1}^n,X^{i-1})\Big]\nonumber\\
	&\; \overset{(c)}{=}\sum_{i=1}^n \Big[I(W;X_i|X^{i-1},Z^{i-1},Y_{i+1}^n) \nonumber\\
	&\qquad\qquad- I(W;Y_i|Y_{i+1}^n,X^{i-1},Z^{i-1})\Big]\nonumber\\
	&\; \overset{(d)}{=}\sum_{i=1}^n \Big[I(W,X^{i-1},Z^{i-1},Y_{i+1}^n;X_i)\nonumber\\
	&\qquad\qquad - I(W,Y_{i+1}^n,X^{i-1},Z^{i-1};Y_i)\Big]\nonumber\\
	& \; \overset{(e)}{=}\sum_{i=1}^n \Big[I(U_i;X_i) - I(U_i;Y_i)\Big]\overset{(f)}{=}\sum_{i=1}^n  I(U_i;X_i|Y_i)\label{eq:privacytoDecconverselossless} 
   \end{align}
	where $(a)$ follows by (\ref{eq:privDec_cons}) and from the Markov chain $W-X^n-Y^n$, $(b)$ follows from Csisz\'{a}r's sum identity \cite{CsiszarKornerbook2011}, $(c)$ follows from the Markov chains
	\begin{align}
		&Z^{i-1}- (X^{i-1},Y_{i+1}^n)-(X_i,W)\label{eq:xizi-1Markov}\\
		&Z^{i-1}- (X^{i-1},Y_{i+1}^n)-(Y_i,W)\label{eq:yizi-1Markov}
	\end{align}
	$(d)$ follows because $X^n$ is i.i.d. and the measurement channels are memoryless, $(e)$ follows from the definition of $U_i$, and $(f)$ follows from the Markov chain $U_i-X_i-Y_i$ for all $i\in[1:n]$.
	
	\textbf{Privacy Leakage to the Eavesdropper}: We obtain
	\begin{align}
	&n(R_{\ell,\text{Eve}}+\delta_n)\nonumber\\ &\overset{(a)}{\geq}[H(W|Z^n)-H(W|Y^n)]+[H(W|Y^n)-H(W|X^n)]\nonumber\\
	&=\sum_{i=1}^n\Big[I(W;Y_i|Y_{i+1}^n)-I(W;Z_i|Z^{i-1})\Big]\nonumber\\
	&\qquad+ \sum_{i=1}^n\Big[I(W;X_i|X^{i-1})-I(W;Y_i|Y_{i+1}^n)\Big]\nonumber\\
	&\overset{(b)}{=}\sum_{i=1}^n\Big[I(W;Y_i|Y_{i+1}^n,Z^{i-1})-I(W;Z_i|Z^{i-1},Y_{i+1}^n)\Big]\nonumber\\
	&\qquad+ \sum_{i=1}^n\Big[I(W;X_i|X^{i-1},Y_{i+1}^n)\!-\!I(W;Y_i|Y_{i+1}^n,X^{i-1})\Big]\nonumber\\
	&\overset{(c)}{=}\sum_{i=1}^n\Big[I(W;Y_i|Y_{i+1}^n,Z^{i-1})-I(W;Z_i|Z^{i-1},Y_{i+1}^n)\Big]\nonumber\\
	&\qquad+ \sum_{i=1}^n\Bigg[I(W;X_i|X^{i-1},Y_{i+1}^n,Z^{i-1})\nonumber\\
	&\qquad\qquad\qquad-\!I(W;Y_i|Y_{i+1}^n,X^{i-1},Z^{i-1})\Bigg]\nonumber
	\end{align}
\begin{align}
	&\overset{(d)}{=}\sum_{i=1}^n\Big[I(W,Y_{i+1}^n,Z^{i-1};Y_i)\!-\!I(W,Z^{i-1},Y_{i+1}^n;Z_i)\Big]\nonumber\\
	&\qquad+ \sum_{i=1}^n\Bigg[I(W,X^{i-1},Y_{i+1}^n,Z^{i-1};X_i)\nonumber\\
	&\qquad\qquad\qquad-I(W,Y_{i+1}^n,X^{i-1},Z^{i-1};Y_i)\Bigg]\nonumber\\
	&\overset{(e)}{=} \sum_{i=1}^n \Big[I(V_i;Y_i)\!-\!I(V_i;Z_i)\!+\!I(U_i,V_i;X_i)\!-\!I(U_i,V_i;Y_i)\Big]\nonumber\\
	&= \sum_{i=1}^n \Bigg[-I(U_i,V_i;Z_i)+I(U_i,V_i;X_i)\nonumber\\
	&\qquad\qquad\qquad+\left(I(U_i;Z_i|V_i)-I(U_i;Y_i|V_i)\right)\Bigg]\nonumber\\
	&\overset{(f)}{\geq}\! \sum_{i=1}^n\Big[I(U_i;X_i|Z_i)\!+\![I(U_i;Z_i|V_i)\!-\!I(U_i;Y_i|V_i)]^-\Big]
	\end{align}
	where $(a)$ follows by (\ref{eq:privEve_cons}) and from the Markov chain $W-X^n-Z^n$, $(b)$ follows from Csisz\'{a}r's sum identity, $(c)$ follows from the Markov chains in (\ref{eq:xizi-1Markov}) and (\ref{eq:yizi-1Markov}), $(d)$ follows because $X^n$ is i.i.d. and the measurement channels are memoryless, $(e)$ follows from the definitions of $V_i$ and $U_i$, and $(f)$ follows from the Markov chain $V_i-U_i-X_i-Z_i$ for all $i\in[1:n]$.

\textbf{Secrecy Leakage (to the Eavesdropper)}: We have
\begin{align}
	&n(R_{\text{s}}+\delta_n)\nonumber\\ &\overset{(a)}{\geq}\![H(W|Z^n)\!-\!H(W|Y^n)]\!+\![H(W|Y^n)\!-\!H(W|\widetilde{X}^n,Y^n)]\nonumber\\
	&\overset{(b)}{=}\sum_{i=1}^n\Big[I(W;Y_i|Y_{i+1}^n)-I(W;Z_i|Z^{i-1})\Big]\nonumber\\
	&\qquad+\Big[ nH(\widetilde{X}|Y)-\sum_{i=1}^nH(\widetilde{X}_i|\widetilde{X}^{i-1},W,Y^n)\Big]\nonumber\\
	&\overset{(c)}{=}\sum_{i=1}^n\Big[I(W;Y_i|Y_{i+1}^n,Z^{i-1})-I(W;Z_i|Z^{i-1},Y_{i+1}^n)\Big]\nonumber\\
	&\qquad+\Big[ nH(\widetilde{X}|Y)-\sum_{i=1}^nH(\widetilde{X}_i|\widetilde{X}^{i-1},W,Y_{i+1}^n,Y_i)\Big]\nonumber\\
	&\overset{(d)}{\geq}\sum_{i=1}^n\Big[I(W,Y_{i+1}^n,Z^{i-1};Y_i)-I(W,Z^{i-1},Y_{i+1}^n;Z_i)\Big]\nonumber\\
	&\qquad+\Big[ nH(\widetilde{X}|Y)-\sum_{i=1}^nH(\widetilde{X}_i|X^{i-1},Z^{i-1},W,Y_{i+1}^n,Y_i)\Big]\nonumber\\
	&\overset{(e)}{=} \sum_{i=1}^n
\Big[I(V_i;Y_i)-I(V_i;Z_i)+I(U_i,V_i;\widetilde{X}_i|Y_i)\Big]\nonumber\\
  &\overset{(f)}{=} \sum_{i=1}^n
\Big[I(V_i;Y_i)\!-\!I(V_i;Z_i)\!+\!I(U_i,V_i;\widetilde{X}_i)\!-\!I(U_i,V_i;Y_i)\Big]\nonumber\\
  &= \sum_{i=1}^n
\Bigg[-\!I(U_i,V_i;Z_i)\!+\!I(U_i,V_i;\widetilde{X}_i)\nonumber\\
&\qquad\qquad\qquad+(I(U_i;Z_i|V_i)\!-\!I(U_i;Y_i|V_i))\Bigg]\nonumber
				\end{align}
\begin{align}
&\overset{(g)}{\geq} \sum_{i=1}^n\Big[ I(U_i;\widetilde{X}_i|Z_i)\!+\![I(U_i;Z_i|V_i)\!-\!I(U_i;Y_i|V_i)]^-\Big]
\end{align}
where $(a)$ follows by (\ref{eq:secrecyleakage_cons}), $(b)$ follows because $(\widetilde{X}^n,Y^n)$ are i.i.d., $(c)$ follows from Csisz\'{a}r's sum identity and the Markov chain in (\ref{eq:markovyiminus1xtildei}), $(d)$ follows because $X^n$ is i.i.d. and the measurement channels are memoryless, and from the data processing inequality applied to the Markov chain in (\ref{eq:Markovconversextildeandximinus1}), $(e)$ follows from the definitions of $V_i$ and $U_i$, $(f)$ follows from the Markov chain $(U_i,V_i)-\widetilde{X}_i-Y_i$ for all $i\in[1:n]$, and $(g)$ follows from the Markov chain $V_i-U_i-\widetilde{X}_i-Z_i$ for all $i\in[1:n]$.
	
Introduce a uniformly distributed time-sharing random variable $\displaystyle Q\!\sim\! \text{Unif}[1\!:\!n]$ independent of other random variables. Define $X\!=\!X_Q$, $\displaystyle \widetilde{X}\!=\!\widetilde{X}_Q$, $\displaystyle Y\!=\!Y_Q$, $\displaystyle Z\!=\!Z_Q$, $V\!=\!V_Q$, $U\!=\!(U_Q,\!Q)$, and $f=f_Q$ so that $\displaystyle (Q,V)\!-U-\widetilde{X}-X-(Y,Z)$ form a Markov chain. The converse proof of Theorem~\ref{theo:innerouter_secrecystorprivregions} follows by letting $\delta_n\rightarrow0$.

\textbf{Cardinality Bounds}: We use the support lemma \cite[Lemma 15.4]{CsiszarKornerbook2011}. One can preserve $P_{\widetilde{X}}$ by using $|\mathcal{\widetilde{X}}|-1$ real-valued continuous functions. We have to preserve two expressions for the two cases such that $I(U;Z|V,Q\!=\!q)\!>\!I(U;Y|V,Q=q)$ and $I(U;Z|V,Q=q)\leq I(U;Y|V,Q=q)$ for all $q\in\mathcal{Q}$, so one can limit the cardinality $|\mathcal{Q}|$ of $Q$ to $|\mathcal{Q}|\leq 2$. Furthermore, we have to preserve five more expressions, i.e., $H(\widetilde{X}|U,V,Z)$, $H(\widetilde{X}|U,V,Y)$, $H(X|U,V,Y)$, $H(X|U,V,Z)$, and $(I(U;Z|V)-I(U;Y|V))$. Thus, one can limit the cardinality $|\mathcal{V}|$ of $V$ to $|\mathcal{V}|\leq|\mathcal{\widetilde{X}}|+4$. Similarly, in addition to the  $|\mathcal{\widetilde{X}}|-1$ real-valued continuous functions, one should preserve the same five expressions for the auxiliary random variable $U$. To satisfy the Markov condition $(Q,V)-U-\widetilde{X}-X-(Y,Z)$, one can limit the cardinality $|\mathcal{U}|$ of $U$ to $|\mathcal{U}|\leq{(|\mathcal{\widetilde{X}}|+4)}^2$.
\end{IEEEproof}

\section{Proof of Theorem~\ref{theo:innerouterlosslesmfc}}\label{sec:proofofTheoremlosslessmfc}\label{sec:proofTheorem3}
\subsection{Achievability (Inner Bound) Proof of Theorem~\ref{theo:innerouterlosslesmfc}}
The achievability proof follows by using the OSRB method, as described below.

\begin{IEEEproof}[Proof Sketch]
	Similar to Section~\ref{subsec:achproofofTheorem1}, fix $P_{U_j|\widetilde{X}_j}$ and $P_{V_j|U_j}$ such that $U_j$ is admissible for the function $f_j(\widetilde{X}_j,Y_j)$ for all $j\in[1:J]$ and let $(V_{[1:J]}^n,U_{[1:J]}^n,\widetilde{X}_{[1:J]}^n,X^n,Y_{[1:J]}^n,Z_{[1:J]}^n)$ be i.i.d. according to (\ref{eq:Theorem3achprobdistribution}). We remark that since all $n$-letter random variables are i.i.d., $U_j^n$ is also admissible for all $j\in[1:J]$. 
	
Assign two random bin indices $(F_{\text{v},j},W_{\text{v},j})$	to each $v_j^n$, and assume $F_{\text{v},j}\in[1:2^{n\widetilde{R}_{\text{v},j}}]$ and $W_{\text{v}, j}\in[1:2^{nR_{\text{v},j}}]$ for all $j\in[1:J]$. Similarly, for all $j\in[1:J]$ assign two indices $(F_{\text{u},j},W_{\text{u},j})$ to each $u_j^n$, where $F_{\text{u},j}\in[1:2^{n\widetilde{R}_{\text{u},j}}]$ and $W_{\text{u},j}\in[1:2^{nR_{\text{u},j}}]$. The public message is $W_j=(W_{\text{v},j}, W_{\text{u},j})$ and indices $F_j=(F_{\text{v},j}, F_{\text{u},j})$ represent the public choice of encoder-decoder pairs for all $j\in[1:J]$.
	
	For all $j\in[1:J]$, using a Slepian-Wolf (SW)  decoder, one can reliably estimate $V_j^n$ from $(F_{\text{v},j},W_{\text{v},j}, Y_j^n)$ if we have 
	\begin{align}
	\widetilde{R}_{\text{v},j} + R_{\text{v},j}> H(V_j|Y_j)\label{eq:Vnreconsmfc}
	\end{align}	
	and one can reliably estimate $U_j^n$ from $(F_{\text{u},j},W_{\text{u},j}, Y_j^n, V_j^n)$ by using a SW decoder if we have
	\begin{align}
	\widetilde{R}_{\text{u},j} + R_{\text{u},j}> H(U_j|V_j,Y_j).\label{eq:Unreconsmfc}
	\end{align}
	Thus, applying the union bound, we can show that the reliability constraint in (\ref{eq:reliability_consmfc}) is satisfied if (\ref{eq:Vnreconsmfc}) and (\ref{eq:Unreconsmfc}) are satisfied for all $j\in[1:J]$.
	
	The public index $F_{\text{v},j}$ is almost independent of $\widetilde{X}_j^n$, so it is almost independent of $(V^n_{[1:J]\setminus\{j\}},U^n_{[1:J]\setminus\{j\}},\widetilde{X}_{[1:J]}^n,X^n,Y_{[1:J]}^n,Z_{[1:J]}^n)$, if we have 
	\begin{align}
	\widetilde{R}_{\text{v},j}<H(V_j|\widetilde{X}_j),\qquad\qquad\forall j\in[1:J].\label{eq:independenceofFvmfc}
	\end{align} 
	The public index $F_{\text{u},j}$ is almost independent of $(V_j^n,\widetilde{X}_j^n)$, so it is almost independent of $(V^n_{[1:J]},U^n_{[1:J]\setminus\{j\}},\widetilde{X}_{[1:J]}^n,X^n,Y_{[1:J]}^n,Z_{[1:J]}^n)$, if we have 
	\begin{align}
	\widetilde{R}_{\text{u},j}<H(U_j|V_j,\widetilde{X}_j),\qquad\qquad\forall j\in[1:J].\label{eq:independenceofFumfc}
	\end{align} 
	
	To satisfy the constraints (\ref{eq:Vnreconsmfc})-(\ref{eq:independenceofFumfc}), similar to Section~\ref{sec:proofofTheorem1}, we fix the rates to
	\begin{alignat}{2}
	&\widetilde{R}_{\text{v},j} \!=\! H(V_j|\widetilde{X}_j)\!-\!\epsilon,&&\;\;\forall j\in [1:J]\label{eq:R_vtildechosenmfc}\\
	&R_{\text{v},j} \!=\! I(V_j;\widetilde{X}_j)\!-\!I(V_j;Y_j)\!+\!2\epsilon,&&\;\;\forall j\in[1:J]\label{eq:R_vchosenmfc}\\
	&\widetilde{R}_{\text{u},j} \!=\! H(U_j|V_j,\widetilde{X}_j)\!-\!\epsilon,&&\;\;\forall j\in[1:J]\label{eq:R_utildechosenmfc}\\
	&R_{\text{u},j} \!=\! I(U_j;\widetilde{X}_j|V_j)\!-\!I(U_j;Y_j|V_j)\!+\!2\epsilon,&&\;\;\forall j\in [1:J]\label{eq:R_uchosenmfc}
	\end{alignat}
	for any $\epsilon>0$. 
	
	\textbf{Storage (Public Message) Rate}: (\ref{eq:R_vtildechosenmfc})-(\ref{eq:R_uchosenmfc}) result in a storage (public message) rate $R_{\text{w},j}$ of
	\begin{align}
	&R_{\text{w},j} = R_{\text{v},j} + R_{\text{u},j}=  I(V_j, U_j;\widetilde{X}_j)-I(V_j,U_j;Y_j)+4\epsilon\nonumber\\
	& \overset{(a)}{=}   I(U_j;\widetilde{X}_j|Y_j)+4\epsilon,\qquad\qquad\qquad\qquad\forall j\in[1:J]\label{eq:R_wchosenmfc}
	\end{align}
	where $(a)$ follows because $V_j-U_j-\widetilde{X}_j-Y_j$ form a Markov chain for all $j\in [1:J]$. 
	
	\textbf{Privacy Leakage to the Decoder}: Consider the privacy leakage to the decoder. We have
	\begin{align}
	&I(X^n;W_j,F_j|Y_j^n) \nonumber\\
	&\qquad\overset{(a)}{\leq}nI(U_j;X|Y_j)+2\epsilon_n,\qquad\qquad\quad\forall j\in [1\!:\!J]\label{eq:ach_privtoDecmfc}
	\end{align}
	where $(a)$ follows for some $\epsilon_n>0$ with $\epsilon_n\rightarrow 0$ when $n\rightarrow\infty$ by applying the steps in (\ref{eq:ach_privtoDec}).
	
	\textbf{Privacy Leakage to the Eavesdropper}: Suppose an additional virtual joint encoder assigns $4J$ indices $(F_{v,[1:J]},W_{v,[1:J]},F_{u,[1:J]},W_{u,[1:J]})$ to each realization tuple
	$(v_1^n,v_2^n,\dots,v^n_{J},u_1^n,u_2^n,\dots,u^n_{J})\in\mathcal{V}_1\times\mathcal{V}_2\times\ldots\times\mathcal{V}_J\times\mathcal{U}_1\times\mathcal{U}_2\times\ldots\times\mathcal{U}_J$ such that 
	\begin{align}
		&\sum_{j=1}^{J}(\widetilde{R}_{\text{v},j} + R_{\text{v},j})>H(V_{[1:J]}|Y_{[1:J]}),\label{eq:virtualcons1}\\
		&\sum_{j=1}^{J}(\widetilde{R}_{\text{u},j} + R_{\text{u},j})> H(U_{[1:J]}|V_{[1:J]},Y_{[1:J]}).\label{eq:virtualcons2}
	\end{align}
	Thus, $(W_{v,[1:J]},F_{v,[1:J]},Y^n_{[1:J]})$ can reliably recover $V^n_{[1:J]}$ and $(V^n_{[1:J]},W_{u,[1:J]},F_{u,[1:J]},Y^n_{[1:J]})$ can reliably recover $U^n_{[1:J]}$. Therefore, we have for the total storage rate that
	\begin{align}
		&\sum_{j=1}^JR_{\text{w},j}=\sum_{j=1}^J(R_{\text{v},j}+R_{\text{u},j})\nonumber\\
		&\qquad \overset{(a)}{\geq}I(U_{[1:J]},V_{[1:J]};\widetilde{X}_{[1:J]})-I(U_{[1:J]},V_{[1:J]};Y_{[1:J]})\nonumber\\
		&\qquad \overset{(b)}{=}I(U_{[1:J]};\widetilde{X}_{[1:J]}|Y_{[1:J]})
	\end{align}
	where $(a)$ follows by (\ref{eq:virtualcons1}) and (\ref{eq:virtualcons2}), and because (\ref{eq:independenceofFvmfc}) and (\ref{eq:independenceofFumfc}) ensure that $(F_{v,[1:J]},F_{u,[1:J]})$ are almost mutually independent of $\widetilde{X}^n_{[1:J]}$ since $\sum_{j=1}^J(\widetilde{R}_{v,j}+\widetilde{R}_{u,j})< H(U_{[1:J]},V_{[1:J]}|\widetilde{X}_{[1:J]})$ such that equality is achieved when $n\rightarrow\infty$ and $(b)$ follows from the Markov chain $V_{[1:J]}- U_{[1:J]}-\widetilde{X}_{[1:J]}-Y_{[1:J]}$. 
	
	Consider the privacy leakage to the eavesdropper. We have
	\begin{align}
	&I(X^n;W_{[1:J]},F_{[1:J]}|Z_{[1:J]}^n)\nonumber\\
	&\overset{(a)}{=}H(W_{[1:J]},F_{[1:J]}|Z_{[1:J]}^n)-H(W_{[1:J]},F_{[1:J]}|X^n)\nonumber\\
    &\overset{(b)}{=}H(W_{[1:J]},F_{[1:J]}|Z_{[1:J]}^n)-nH(U_{[1:J]},V_{[1:J]}|X)\nonumber\\
	&\quad+\sum_{j=1}^J \Big[H(V^n_j|V^n_{[1:j-1]},W_{[1:J]},F_{[1:J]},X^n)\nonumber\\
	&\qquad\qquad\quad+H(U^n_j|U^n_{[1:j-1]},V^n_{[1:J]},W_{[1:J]},F_{[1:J]},X^n)\Big]\nonumber\\
	&\overset{(c)}{\leq}\!H(W_{[1:J]},F_{[1:J]}|Z_{[1:J]}^n)\nonumber\\
	&\qquad-nH(U_{[1:J]},V_{[1:J]}|X)+2Jn\epsilon_n^{\prime} \label{eq:ach_privtoEvefirststepmfc}
	\end{align}
	where $(a)$ follows from the Markov chain $Z_{[1:J]}^n-X^n-(W_{[1:J]},F_{[1:J]})$, $(b)$ follows since $U_j^n$ determines $(W_{u,j},F_{u,j})$ and $V_j^n$ determines $(W_{v,j},F_{v,j})$ for all $j\in [1:J]$, and $(U^n_{[1:J]},V^n_{[1:J]},X^n)$ are i.i.d., and $(c)$ follows for some $\epsilon_n^{\prime}>0$ such that $\epsilon^{\prime}_n\rightarrow 0$ when $n\rightarrow\infty$ because $(F_{v,j},W_{v,j}X^n)$ can reliably recover $V_j^n$ due to the Markov chain $V_j^n-X^n-Y_j^n$ and (\ref{eq:Vnreconsmfc}), and because $(W_{u,j},F_{u,j},V_j^n,X^n)$ can reliably recover $U_j^n$ due to the  inequality $H(U_j|V_j,Y_j)\geq H(U_j|V_j,X)$, proved in (\ref{eq:proofthatHUVYisgreaterthanHUVX}), for all $j\in [1:J]$.
	  
	 We consider the six decodability cases considered in Section~\ref{subsec:achproofofTheorem1} by replacing $[(R_v+\widetilde{R}_v),(R_u+\widetilde{R}_u)]$ with $\Big[\Big(\sum_{j=1}^J(R_{v,j}+\widetilde{R}_{v,j})\Big),\Big(\sum_{j=1}^J(R_{u,j}+\widetilde{R}_{u,j})\Big)\Big]$, respectively, and $[H(V|Z),H(U|V,Z), H(U|Z)]$ with $[H(V_{[1:J]}|Z_{[1:J]}),H(U_{[1:J]}|V_{[1:J]},Z_{[1:J]}), H(U_{[1:J]}|Z_{[1:J]})]$, respectively. Using these replacements, applying the steps in (\ref{eq:ach_privtoEveCase1}),(\ref{eq:ach_privtoEveCase2}), (\ref{eq:ach_privtoEveCase3}), (\ref{eq:ach_privtoEveCase4}), (\ref{eq:ach_privtoEveCase5}), and (\ref{eq:ach_privtoEveCase6}) in combination with (\ref{eq:ach_privtoEvefirststepmfc}), and by choosing trivial rates that satisfy (\ref{eq:virtualcons1}) and (\ref{eq:virtualcons2}), one can show that
	  \begin{align}
			&I(X^n;W_{[1:J]},F_{[1:J]}|Z_{[1:J]}^n)\nonumber\\
			&\leq n[I(U_{[1:J]};Z_{[1:J]}|V_{[1:J]})\!-\!I(U_{[1:J]};Y_{[1:J]}|V_{[1:J]})\!+\!\epsilon]^-\nonumber\\
			&\qquad+n(I(U_{[1:J]};X|Z_{[1:J]})+3J\epsilon^{\prime}_n).\label{eq:privacytoEvemfc}
	  \end{align}

	\textbf{Secrecy Leakage (to the Eavesdropper)}: Consider the secrecy leakage. We have
	\begin{align}
	&I(\widetilde{X}_{[1:J]}^n,Y_{[1:J]}^n;W_{[1:J]},F_{[1:J]}|Z_{[1:J]}^n)\nonumber\\
	&\!\overset{(a)}{=}H(W_{[1:J]},F_{[1:J]}|Z_{[1:J]}^n)-H(W_{[1:J]},F_{[1:J]}|\widetilde{X}_{[1:J]}^n)\nonumber\\
	&\! \overset{(b)}{\leq}\! H(W_{[1:J]},F_{[1:J]}|Z_{[1:J]}^n)\nonumber\\
	&\qquad - H(U^n_{[1:J]},V^n_{[1:J]}|\widetilde{X}^n_{[1:J]})\!+\!2Jn\epsilon^{\prime}_n
	\label{eq:ach_secrecyfirststepmfc}
	\end{align}
	where $(a)$ follows from the Markov chain $(W_{[1:J]},F_{[1:J]})-\widetilde{X}_{[1:J]}^n-(Y_{[1:J]}^n,Z_{[1:J]}^n)$, $(b)$ follows for some $\epsilon_n^{\prime}>0$ such that $\epsilon_n^{\prime}\rightarrow 0$ when $n\rightarrow\infty$ because $U^n_j$ determines $(W_{u,j},F_{u,j})$ and $V^n_j$ determines $(W_{v,j},F_{v,j})$, and $(W_{v,j},F_{v,j},\widetilde{X}_j^n)$ can reliably recover $V_j^n$ due to the Markov chain $V_j^n-\widetilde{X}_j^n-Y_j^n$ and (\ref{eq:Vnreconsmfc}), and similarly $(W_{u,j},F_{u,j},V_j^n,\widetilde{X}_j^n)$ can reliably recover $U_j^n$ because $H(U_j|V_j,Y_j)\geq H(U_j|V_j,\widetilde{X}_j)$, which can be proved as in (\ref{eq:proofthatHUVYisgreaterthanHUVX}).
	
	By using the same joint virtual encoder used for the privacy-leakage to the eavesdropper analysis above and replacing $X$ by $\widetilde{X}_{[1:J]}$ in the analyses of (\ref{eq:privacytoEvemfc}), we obtain  from (\ref{eq:ach_secrecyfirststepmfc}) that
	\begin{align}
	&I(\widetilde{X}_{[1:J]}^n,Y_{[1:J]}^n;W_{[1:J]},F_{[1:J]}|Z_{[1:J]}^n)\nonumber\\
	&\!\leq n[I(U_{[1:J]};Z_{[1:J]}|V_{[1:J]})\!-\!I(U_{[1:J]};Y_{[1:J]}|V_{[1:J]})\!+\!\epsilon]^-\!\nonumber\\
	&\qquad +n(I(U_{[1:J]};\widetilde{X}_{[1:J]}|Z_{[1:J]})+\!\!3J\epsilon^{\prime}_n)\label{eq:ach_secrecyleakmfc}.
	\end{align}
	
	Suppose the public indices $F_{[1:J]}$ are generated uniformly at random. The encoder $\Enc_j(\cdot)$ generates $(V_j^n,U_j^n)$ according to $P_{V_j^nU_j^n|\widetilde{X}_j^nF_{\text{v},j}F_{\text{u},j}}$ obtained from the binning scheme above to compute the bins $W_{\text{v},j}$ from $V_j^n$ and $W_{\text{u},j}$ from $U^n_j$, respectively, for all $j\in [1:J]$. This procedure induces a joint probability distribution that is almost equal to $P_{V_{[1:J]}U_{[1:J]}\widetilde{X}_{[1:J]}XY_{[1:J]}Z_{[1:J]}}$ fixed above \cite[Section 1.6]{BlochLectureNotes2018}. We remark that the privacy and secrecy leakage metrics considered above are expectations over all possible realizations $F_{[1:J]}=f_{[1:J]}$. Thus, applying the selection lemma to each decodability case separately, these results prove the achievability for the rate tuples given in Theorem~\ref{theo:innerouterlosslesmfc} by choosing an $\epsilon>0$ such that $\epsilon\rightarrow 0$ when $n\rightarrow\infty$.
\end{IEEEproof}
	\subsection{Converse (Outer Bound) Proof of Theorem~\ref{theo:innerouterlosslesmfc}}
	\begin{IEEEproof}[Proof Sketch]
		Suppose for some $\delta_n\!>\!0$ and $n\geq 1$, there exists a pair of encoders and decoders such that (\ref{eq:reliability_consmfc})-(\ref{eq:privEve_consmfc}) are satisfied for some tuple $(R_\text{s}, R_{\text{w},[1:J]},R_{\ell,{\text{Dec}},[1:J]}, R_{\ell,{\text{Eve}}})$ .  
		
		Let $V_{i,j}\triangleq (W_j,Y^{n}_{i+1,j},Z_j^{i-1})$ and $U_{i,j}\triangleq (W_j,X^{i-1},Y^{n}_{i+1,j},Z_j^{i-1})$, which satisfy the Markov chain $V_{i,j}-U_{i,j}-\widetilde{X}_{i,j}-X_i-(Y_{i,j},Z_{i,j})$ for all $i\in[1:n]$ and $j\in[1:J]$ by definition of the source statistics.

		\textbf{Admissibility of $\mathbf{U_j}$}: Define $\displaystyle \epsilon_n\!=\max_{j\in[1:J]}\!\Big(\delta_{n,j} |\mathcal{\widetilde{X}}_j||\mathcal{Y}_j| \!+\!H_b(\delta_{n,j})/n\Big)$ so that $\epsilon_n\!\rightarrow\!0$ if $\displaystyle\max_{j\in[1:J]}\delta_{n,j}=\delta_n\!\rightarrow\!0$. Applying the union bound to (\ref{eq:reliability_consmfc}) and using Fano's inequality, we obtain
		\begin{align}
		&n\epsilon_n\!\geq\! H(f_j^n|\widehat{f_j^n})\!\overset{(a)}{\geq}\!\sum_{i=1}^n\!\!H(f_{i,j}|U_{i,j},Y_{i,j}),\; \forall j\in[1:J]\label{eq:fanoappmfc} 
		\end{align}
		where $(a)$ follows applying the steps in (\ref{eq:fanoapp}) and from the definition of $U_{i,j}$.
		
		\textbf{Storage (Public Message) Rate}: We have for all $j\in[1:J]$ that
		\begin{align}
		&n(R_{\text{w},j}\!+\!\delta_n)\! \overset{(a)}{\geq} \!\log|\mathcal{W}_j|\!\overset{(b)}{\geq}\!\sum_{i=1}^n I(U_{i,j};\widetilde{X}_{i,j}|Y_{i,j})
		\end{align}
		where $(a)$ follows by (\ref{eq:storage_consmfc}) and $(b)$ follows by applying the steps in (\ref{eq:storagerateconv}) and from the definition of $U_{i,j}$.

	    \textbf{Privacy Leakage to the Decoder}: We obtain for all $j\in[1:J]$ that
		\begin{align}
		&n(R_{\ell,\text{Dec},j}+\delta_n) \overset{(a)}{\geq}H(W_j|Y_j^n)-H(W_j|X^n)\nonumber\\
		&\quad\overset{(b)}{\geq}\sum_{i=1}^n  I(U_{i,j};X_i|Y_{i,j}) 
		\end{align}
		where $(a)$ follows by (\ref{eq:privDec_consmfc}) and from the Markov chain $W_j-X^n-Y_j^n$ and $(b)$ follows by applying the steps in (\ref{eq:privacytoDecconverselossless}) and from the definition of $U_{i,j}$.

				\textbf{Sum-Storage Rate}: We have for all $j\in[1:J]$ that
		\begin{align}
		&n\sum_{j=1}^J(R_{\text{w},j}\!+\!\delta_n)\! \overset{(a)}{\geq} \log\Bigg|\prod_{j=1}^J|\mathcal{W}_j|\Bigg|\nonumber\\
		&\geq\!H(W_{[1:J]}|Y^n_{[1:J]})-H(W_{[1:J]}|\widetilde{X}^n_{[1:J]},Y^n_{[1:J]})\nonumber\\
		&= H(\widetilde{X}^n_{[1:J]}|Y^n_{[1:J]}) - \sum_{i=1}^nH(\widetilde{X}_{i,[1:J]}|\widetilde{X}^{i-1}_{[1:J]},Y^n_{[1:J]}, W_{[1:J]})\nonumber\\
		&\overset{(b)}{=}H(\widetilde{X}^n_{[1:J]}|Y^n_{[1:J]})\nonumber\\
		&\qquad - \sum_{i=1}^nH(\widetilde{X}_{i,[1:J]}|\widetilde{X}^{i-1}_{[1:J]},Y^n_{i+1,[1:J]}, Y_{i,[1:J]},W_{[1:J]})\nonumber\\
		&\overset{(c)}{\geq}H(\widetilde{X}^n_{[1:J]}|Y^n_{[1:J]})\nonumber\\
		&\qquad - \sum_{i=1}^nH(\widetilde{X}_{i,[1:J]}|X^{i-1}_{[1:J]},Z^{i-1}_{[1:J]},Y^n_{i+1,[1:J]},Y_{i,[1:J]},W_{[1:J]})\nonumber\\
		&\overset{(d)}{=}\sum_{i=1}^{n}I(U_{i,[1:J]};\widetilde{X}_{i,[1:J]}|Y_{i,[1:J]})
		\end{align}
		where $(a)$ follows by (\ref{eq:storage_consmfc}), $(b)$ follows from the Markov chain  for all $i\in [1:n]$
		\begin{align}
		Y_{[1:J]}^{i-1}-(\widetilde{X}_{[1:J]}^{i-1},W_{[1:J]},Y_{i,[1:J]}^n)-\widetilde{X}_{i,[1:J]}\label{eq:markovyiminus1xtildeimfc}
		\end{align}
		$(c)$ follows from applying the data processing inequality to the Markov chain for all $i\in[1:n]$
		\begin{align}
		&(X^{i-1},Z_{[1:J]}^{i-1})-(\widetilde{X}_{[1:J]}^{i-1}, W_{[1:J]},Y_{i,[1:J]}^n)-\widetilde{X}_{i,[1:J]}\label{eq:Markovconversextildeandximinus1mfc}
		\end{align}
		and $(d)$ follows because $(\widetilde{X}^n_{[1:J]},Y^n_{[1:J]})$ are i.i.d. and from the definition of $U_{i,j}$ for all $j\in [1:J]$.

		\textbf{Privacy Leakage to the Eavesdropper}: We obtain
		\begin{align}
		&n(R_{\ell,\text{Eve}}+\delta_n)\nonumber\\ &\overset{(a)}{\geq}[H(W_{[1:J]}|Z_{[1:J]}^n)-H(W_{[1:J]}|Y_{[1:J]}^n)]\nonumber\\
		&\qquad +[H(W_{[1:J]}|Y_{[1:J]}^n)-H(W_{[1:J]}|X^n)]\nonumber
													\end{align}
		\begin{align}
		&\overset{(b)}{=}\sum_{i=1}^n\Big[I(W_{[1:J]};Y_{i,[1:J]}|Y_{i+1,[1:J]}^n,Z_{[1:J]}^{i-1})\nonumber\\
		&\qquad\qquad-I(W_{[1:J]};Z_{i,[1:J}]|Z_{[1:J]}^{i-1},Y_{i+1,[1:J]}^n)\Big]\nonumber\\
		&\qquad+ \sum_{i=1}^n\Big[I(W_{[1:J]};X_i|X^{i-1},Y_{i+1,[1:J]}^n)\nonumber\\
		&\qquad\qquad\quad\;-\!I(W_{[1:J]};Y_{i,[1:J]}|Y_{i+1,[1:J]}^n,X^{i-1})\Big]\nonumber\\
		&\overset{(c)}{=}\sum_{i=1}^n\Big[I(W_{[1:J]};Y_{i,[1:J]}|Y_{i+1,[1:J]}^n,Z_{[1:J]}^{i-1})\nonumber\\
		&\qquad\qquad-I(W_{[1:J]};Z_{i,[1:J}]|Z_{[1:J]}^{i-1},Y_{i+1,[1:J]}^n)\Big]\nonumber\\
		&\qquad+ \sum_{i=1}^n\Big[I(W_{[1:J]};X_i|X^{i-1},Y_{i+1,[1:J]}^n,Z^{i-1}_{[1:J]})\nonumber\\
		&\qquad\qquad\quad\;-\!I(W_{[1:J]};Y_{i,[1:J]}|Y_{i+1,[1:J]}^n,X^{i-1},Z^{i-1}_{[1:J]})\Big]\nonumber\\
		&\overset{(d)}{=}\sum_{i=1}^n\Big[I(W_{[1:J]},Y_{i+1,[1:J]}^n,Z^{i-1}_{[1:J]};Y_{i,[1:J]})\nonumber\\
		&\qquad\qquad-\!I(W_{[1:J]},Z_{[1:J]}^{i-1},Y_{i+1,[1:J]}^n;Z_{i,[1:J]})\Big]\nonumber\\
		&\qquad+ \sum_{i=1}^n\Big[I(W_{[1:J]},X^{i-1},Y_{i+1,[1:J]}^n,Z_{[1:J]}^{i-1};X_i)\nonumber\\
		&\qquad\qquad\quad\;-I(W_{[1:J]},Y_{i+1,[1:J]}^n,X^{i-1},Z_{[1:J]}^{i-1};Y_{i,[1:J]})\Big]\nonumber\\
		&\overset{(e)}{=} \sum_{i=1}^n \Big[I(V_{i,[1:J]};Y_{i,[1:J]})-I(V_{i,[1:J]};Z_{i,[1:J]})\nonumber\\
		&\qquad\qquad\!+\!I(U_{i,[1:J]},V_{i,[1:J]};X_i)\!-\!I(U_{i,[1:J]},V_{i,[1:J]};Y_{i,[1:J]})\Big]\nonumber\\
		&\!=\! \sum_{i=1}^n \Bigg[\!-\!I(U_{i,[1:J]},V_{i,[1:J]};Z_{i,[1:J]})\!+\!I(U_{i,[1:J]},V_{i,[1:J]};X_{i})\nonumber\\
		&\qquad\qquad+I(U_{i,[1:J]};Z_{i,[1:J]}|V_{i,[1:J]})\nonumber\\
		&\qquad\qquad-I(U_{i,[1:J]};Y_{i,[1:J]}|V_{i,[1:J]})\Bigg]\nonumber\\
		&\!\overset{(f)}{\geq}\!\! \sum_{i=1}^n\!\Bigg[\!\big[\!I(U_{i,[1:J]};Z_{i,[1:J]}|V_{i,[1:J]})\!-\!I(U_{i,[1:J]};Y_{i,[1:J]}|V_{i,[1:J]})\big]^-\nonumber\\
		&\qquad\qquad+I(U_{i,[1:J]};X_i|Z_{i,[1:J]})\Bigg]
		\end{align}
		where $(a)$ follows by (\ref{eq:privEve_consmfc}) and from the Markov chain $W_{[1:J]}-X^n-Z_{[1:J]}^n$, $(b)$ follows from Csisz\'{a}r's sum identity, $(c)$ follows from the Markov chains for all $i\in [1:n]$
		\begin{align}
		&Z_{[1:J]}^{i-1}- (X^{i-1},Y_{i+1,[1:J]}^n)-(X_i,W_{[1:J]})\label{eq:xizi-1Markovmfc}\\
		&Z_{[1:J]}^{i-1}- (X^{i-1},Y_{i+1,[1:J]}^n)-(Y_{i,[1:J]},W_{[1:J]})\label{eq:yizi-1Markovmfc}
		\end{align}
		 $(d)$ follows because $X^n$ is i.i.d. and the measurement channels are memoryless, $(e)$ follows from the definitions of $V_{i,j}$ and $U_{i,j}$ for all $j\in[1:J]$, and $(f)$ follows from the Markov chain $V_{i,[1:J]}-U_{i,[1:J]}-X_i-Z_{i,[1:J]}$ for all $i\in[1:n]$.

		\textbf{Secrecy Leakage (to the Eavesdropper)}: We have
		\begin{align}
		&n(R_{\text{s}}+\delta_n)\nonumber\\ &\overset{(a)}{\geq}\![H(W_{[1:J]}|Z_{[1:J]}^n)\!-\!H(W_{[1:J]}|Y_{[1:J]}^n)]\nonumber\\
		&\qquad+\![H(W_{[1:J]}|Y_{[1:J]}^n)\!-\!H(W_{[1:J]}|\widetilde{X}_{[1:J]}^n,Y_{[1:J]}^n)]\nonumber\\
		&\overset{(b)}{=}\sum_{i=1}^n\Big[I(W_{[1:J]};Y_{i,[1:J]}|Y_{i+1,[1:J]}^n)\nonumber\\
		&\qquad\qquad-\!I(W_{[1:J]};Z_{i,[1:J]}|Z_{[1:J]}^{i-1})\Big]\nonumber\\
		&\qquad+\Big[ nH(\widetilde{X}_{[1:J]}|Y_{[1:J]})\nonumber\\
		&\qquad\qquad-\!\sum_{i=1}^nH(\widetilde{X}_{i,[1:J]}|\widetilde{X}_{[1:J]}^{i-1},W_{[1:J]},Y_{[1:J]}^n)\Big]\nonumber\\
		&\overset{(c)}{=}\sum_{i=1}^n\Big[I(W_{[1:J]};Y_{i,[1:J]}|Y_{i+1,[1:J]}^n,Z_{[1:J]}^{i-1})\nonumber\\
		&\qquad\qquad-I(W_{[1:J]};Z_{i,[1:J]}|Z_{[1:J]}^{i-1},Y_{i+1,[1:J]}^n)\Big]\nonumber\\
		&\qquad+\Big[ nH(\widetilde{X}_{[1:J]}|Y_{[1:J]})\nonumber\\
		&\qquad\qquad-\sum_{i=1}^nH(\widetilde{X}_{i,[1:J]}|\widetilde{X}_{[1:J]}^{i-1},W_{[1:J]},Y_{i+1,[1:J]}^n,Y_{i,[1:J]})\Big]\nonumber\\
		&\overset{(d)}{\geq}\sum_{i=1}^n\Big[I(W_{[1:J]},Y_{i+1,[1:J]}^n,Z_{[1:J]}^{i-1};Y_{i,[1:J]})\nonumber\\
		&\qquad\qquad-I(W_{[1:J]},Z_{[1:J]}^{i-1},Y_{i+1,[1:J]}^n;Z_{i,[1:J]})\Big]\nonumber\\
		&\qquad+\Big[ nH(\widetilde{X}_{[1:J]}|Y_{[1:J]})\nonumber\\
		&\qquad\quad\!-\!\sum_{i=1}^n\!H(\widetilde{X}_{i,[1\!:J]}|X^{i-1},Z_{[1\!:J]}^{i-1},W_{[1\!:J]},Y_{i+1,[1\!:J]}^n,Y_{i,[1\!:J]})\Big]\nonumber\\
		&\overset{(e)}{=} \sum_{i=1}^n
		\Big[I(V_{i,[1:J]};Y_{i,[1:J]})-I(V_{i,[1:J]};Z_{i,[1:J]})\nonumber\\
		&\qquad\qquad+I(U_{i,[1:J]},V_{i,[1:J]};\widetilde{X}_{i,[1:J]}|Y_{i,[1:J]})\Big]\nonumber\\
		&\overset{(f)}{=} \sum_{i=1}^n
		\Big[I(V_{i,[1:J]};Y_{i,[1:J]})\!-\!I(V_{i,[1:J]};Z_{i,[1:J]})\nonumber\\
		&\qquad\qquad+\!I(U_{i,[1:J]},V_{i,[1:J]};\widetilde{X}_{i,[1:J]})\nonumber\\
		&\qquad\qquad-\!I(U_{i,[1:J]},V_{i,[1:J]};Y_{i,[1:J]})\Big]\nonumber\\
		&= \sum_{i=1}^n
		\!\Big[\!-\!I(U_{i,[1:J]},V_{i,[1:J]};Z_{i,[1:J]})\nonumber\\
		&\qquad\qquad+\!I(U_{i,[1:J]},V_{i,[1:J]};\widetilde{X}_{i,[1:J]})\nonumber\\
		&\qquad\qquad+I(U_{i,[1:J]};Z_{i,[1:J]}|V_{i,[1:J]})\nonumber\\
		&\qquad\qquad-\!I(U_{i,[1:J]};Y_{i,[1:J]}|V_{i,[1:J]})\Big]\nonumber\\
		&\!\overset{(g)}{\geq}\! \sum_{i=1}^n\!\Bigg[\![I(U_{i,[1:J]};Z_{i,[1:J]}|V_{i,[1:J}])\!-\!I(U_{i,[1:J]};Y_{i,[1:J]}|V_{i,[1:J]})]^-\!\nonumber\\
		&\qquad\qquad  +I(U_{i,[1:J]};\widetilde{X}_{i,[1:J]}|Z_{i,[1:J]})\Bigg]
		\end{align}
		where $(a)$ follows by (\ref{eq:secrecyleakage_consmfc}), $(b)$ follows since $(\widetilde{X}_{[1:J]}^n,Y_{[1:J]}^n)$ are i.i.d., $(c)$ follows from Csisz\'{a}r's sum identity and the Markov chain  in (\ref{eq:markovyiminus1xtildeimfc}), $(d)$ follows because $X^n$ is i.i.d. and the measurement channels are memoryless, and from the data processing inequality applied to the Markov chain in (\ref{eq:Markovconversextildeandximinus1mfc}),  $(e)$ follows from the definitions of $V_{i,[1:J]}$ and $U_{i,[1:J]}$, $(f)$ follows from the Markov chain $(U_{i,[1:J]},V_{i,[1:J]})-\widetilde{X}_{i,[1:J]}-Y_{i,[1:J]}$ for all $i\in[1:n]$, and $(g)$ follows from the Markov chain $V_{i,[1:J]}-U_{i,[1:J]}-\widetilde{X}_{i,[1:J]}-Z_{i,[1:J]}$ for all $i\in[1:n]$.
		
		Introduce a uniformly distributed time-sharing random variable $\displaystyle Q\!\sim\! \text{Unif}[1\!:\!n]$ independent of other random variables. Define $X\!=\!X_Q$, $\displaystyle \widetilde{X}_j\!=\!\widetilde{X}_{Q,j}$, $\displaystyle Y_j\!=\!Y_{Q,j}$, $\displaystyle Z_j\!=\!Z_{Q,j}$, $V_j\!=\!V_{Q,j}$, $U_j\!=\!(U_{Q,j},\!Q)$, and $f_j=f_{Q,j}$ so that $\displaystyle (Q,V_j)\!-U_j-\widetilde{X}_j-X-(Y_j,Z_j)$ form a Markov chain for all $j\in[1:J]$. The converse proof of Theorem~\ref{theo:innerouterlosslesmfc} follows by letting $\delta_n\rightarrow0$.
		
		\textbf{Cardinality Bounds} follow by using the support lemma as in Section~\ref{subsec:converseofTheorem1}.
\end{IEEEproof}

\section{Conclusion}\label{sec:conclusion}
We derived the secrecy-storage-privacyDec-privacyEve(-distortion) regions for lossless and lossy single-function computations with a remote source. The remote source model allows to model multiple sequences observed by a single terminal as multiple noisy measurements of a hidden source, which allows to measure the diversity gains. The equivocation measure common in the literature was replaced with a mutual information metric, which resulted in simpler notation and easier interpretations. A new privacy metric was considered to bound the information leakage to a fusion center about the remote source sequence. Bounds for the storage and privacy leakage to the eavesdropper rates were shown to be different, unlike in the previous models. Inner and outer bounds for multiple asynchronous function computations within the same network were given to illustrate the effects of joint constraints for all terminals involved in any function computation. These bounds differ only in the Markov chain conditions imposed. We evaluated the rate region for a single-function computation problem by solving an information bottleneck problem for binary input symmetric output channels. In future work, we will consider multi-function computations with multiple transmitting terminals for each function computation and derive the rate regions for two-function computations with two transmitting terminals if a set of symmetry conditions are satisfied.

\IEEEtriggeratref{39}
\bibliographystyle{IEEEtran}
\bibliography{references}

\end{document}